\documentclass[tighten,nofootinbib,amssymb,floatfix]{revtex4}
\usepackage{bm}% bold math
\usepackage{graphicx}%
\usepackage{epsf}
\usepackage{epsfig}
\usepackage{color,bm}

\usepackage{ulem}

\usepackage{amsmath}
\usepackage{amsfonts,amssymb}
\newcommand{\tr}{\mbox{Tr}}                      % Trace symbol
\newcommand{\re}{\operatorname{Re}}

\newcommand{\Op}{\mathcal{O}} % Fractur O
\newcommand{\eins}{\mathds{1}} % Operator 1
\newcommand{\Dlr}{\buildrel \leftrightarrow \over D\raise-1pt\hbox{}}
\newcommand{\Dl}{\buildrel \leftarrow \over D\raise-1pt\hbox{}}
\newcommand{\Dr}{\buildrel \rightarrow \over D\raise-1pt\hbox{}}
  
\newcommand{\ZV}{Z_{\rm V}}                      % \ZV
\def\Zq{Z_{\rm q}}
\def\ZS{Z_{\rm S}}
\def\ZP{Z_{\rm P}}
\def\ZV{Z_{\rm V}}
\def\ZA{Z_{\rm A}}
\def\ZT{Z_{\rm T}}

\def\ZDVa{Z_{\rm DV1}}
\def\ZDVb{Z_{\rm DV2}}
\def\ZDAa{Z_{\rm DA1}}
\def\ZDAb{Z_{\rm DA2}}
\def\ZDTa{Z_{\rm DT1}}
\def\ZDTb{Z_{\rm DT2}}
\newcommand{\J}{\mathcal{J}} % operator
\newcommand{\s}{{\mathcal S}} % Charge conjugation matrix

\newcommand{\be}{\begin{equation}}
\newcommand{\ee}{\end{equation}}
\newcommand{\bea}{\begin{eqnarray}}
\newcommand{\eea}{\end{eqnarray}}
\newcommand{\csw}{\, c_{\rm sw}}

\def\smn{{\sigma_{\mu\nu}}}

\usepackage{dsfont}
\usepackage{slashed}
\usepackage{booktabs}
\usepackage{placeins}
\usepackage{wasysym}

\begin{document}

\title{Renormalization functions for $N_f{=}2$ and $N_f{=}4$ Twisted Mass fermions

\vskip 1cm
\centerline{\psfig{figure=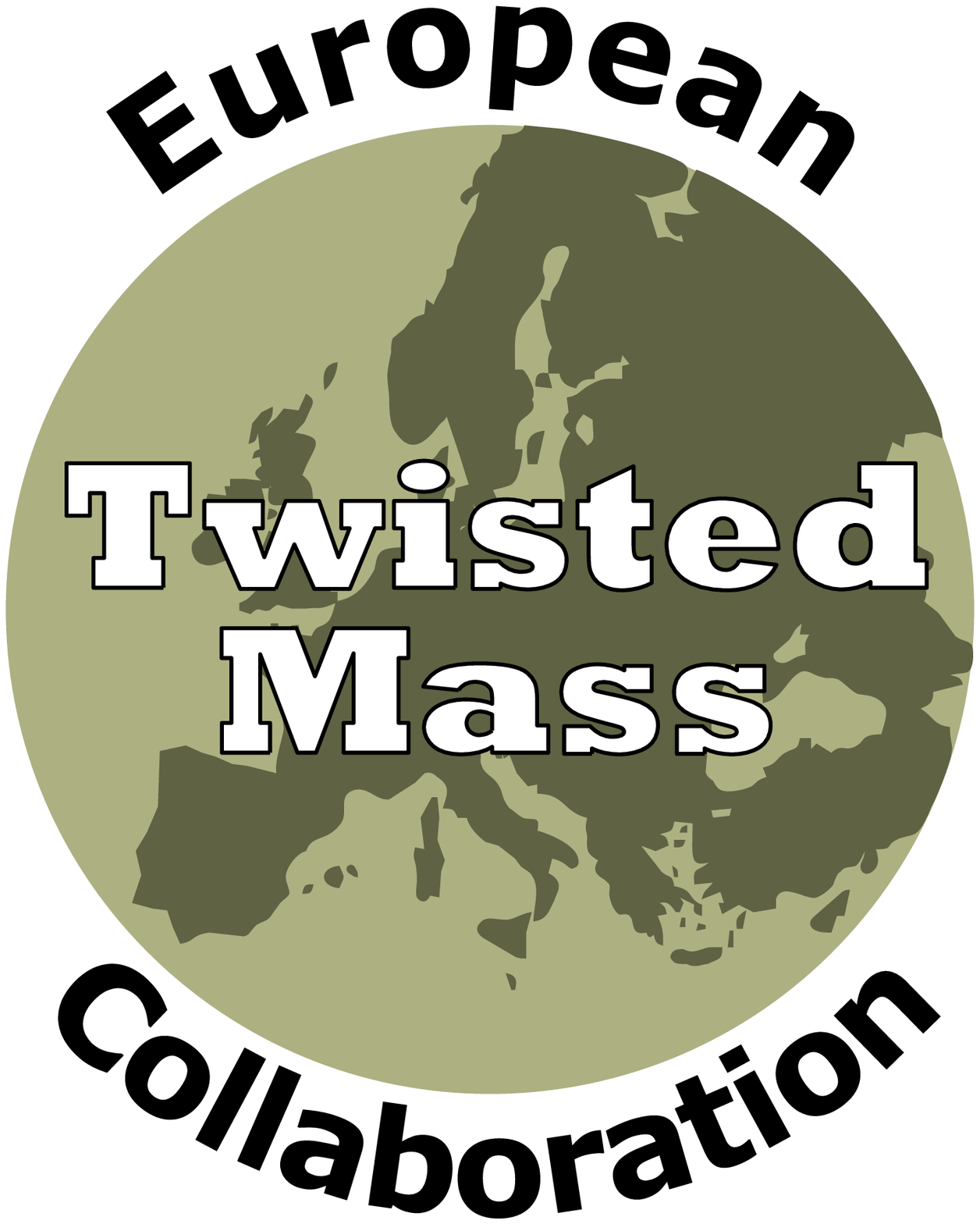,height=2.2truecm}}}

\vskip 1cm

\author{C. Alexandrou$^{a, b}$, M.~Constantinou$^{a, b}$, H.~Panagopoulos$^a\,$}
\email{alexand@ucy.ac.cy,   m.constantinou@cyi.ac.cy,  haris@ucy.ac.cy}

\affiliation{\\ \vskip 0.35cm
$a\,\, Department\,\, of\,\, Physics,\,\, University\,\, of\,\, Cyprus,$\\ 
$POB\,\, 20537,\,\, 1678\,\, Nicosia,\,\, Cyprus$   \\ \vskip 0.25cm
$b\,\, Computation\,\, based\,\, Science\,\, and\,\, Technology\,\, Research\,\, Center,\,\, 
The\,\, Cyprus\,\, Institute,$ \\
$15\,\, Kypranoros\,\, Str.,\,\, 1645\,\, Nicosia,\,\, Cyprus$\\\vskip 0.25cm
}

\begin{abstract} 
We present results on the
renormalization functions of the quark field and fermion bilinears
with up to one covariant derivative. For the fermion part of the
action we employ the twisted mass formulation with $N_f{=}2$ and $N_f{=}4$
degenerate dynamical quarks, while in the gluon sector we use the Iwasaki
improved action. The simulations for $N_f{=}4$ have been performed for
pion masses in the range of 390~MeV - 760~MeV and at three values of the
lattice spacing, $a$, corresponding to $\beta{=}1.90,\,1.95,\,2.10$. 
The $N_f{=}2$ action includes a clover term with $\csw{=}1.57551$ at
$\beta{=}2.10$, and three ensembles at different values of $m_\pi$.

The evaluation of the renormalization functions is carried out in the RI$'$
scheme using a momentum source. The non-perturbartive evaluation is
complemented with a
perturbative computation, which is carried out at one-loop level and to
all orders in the lattice spacing, $a$. For each renormalization function
computed non-perturbatively we subtract the corresponding lattice
artifacts to all orders in $a$, so that  a large part of
the cut-off effects is eliminated.  

The renormalization functions are converted to the ${\overline{\rm MS}}$ 
scheme at a reference energy scale of $\mu{=}2$ GeV after taking the chiral limit. 

\end{abstract}
\pacs{11.15.Ha, 12.38.Gc, 12.38.Aw, 12.38.-t, 14.70.Dj}
\keywords{Lattice QCD, Twisted mass fermions, Renormalization functions, Improvement}

\maketitle
\newpage
\section{Introduction}

Over the last years, simulations of Quantum Chromodynamics (QCD) have
advanced remarkably and are, nowadays, being carried out at
close-to-physical values for the parameters of the theory. Therefore,
{\it ab initio} calculations of hadron structure within lattice QCD
yield results that can be connected to experiment more reliably than
ever before. 
A number of lattice groups are producing results on nucleon form
factors and first moments of structure functions at or close to the
physical regime both in terms of pion mass as well as in terms of the
continuum limit (see Ref.~\cite{Constantinou:2014tga} and references
therein).  At the same time properties of other hadrons that are
difficult to study  experimentally are being pursued within lattice
QCD. These include the axial charges of resonances such as the
$\Delta$~\cite{Alexandrou:2011py} or other nucleon excited
states~\cite{Takahashi:2008fy},
hyperons~\cite{Lin:2007ap,Erkol:2009ev,Alexandrou:2014vya} and charm
baryons~\cite{Alexandrou:2014vya}. For all these quantities, one needs the
renormalization functions in  order to obtain the continuum
predictions. Moments of generalized parton distributions (GPDs) are
connected to generalized form factors and  provide detailed
information on the internal structure of hadrons in terms of both the
longitudinal momentum fraction and the total momentum transfer
squared. Beyond the information that the form factors yield, such as
size, magnetization and shape, GPDs encode additional information,
relevant for experimental investigations, such as the decomposition of
the total hadron spin into angular momentum and spin carried by quarks
and gluons. In lattice QCD one  calculates matrix elements of fermion
operators between the relevant  hadron states and unless these
operators correspond to a conserved current they must be renormalized
in order to extract the  physical information one is after.
In many cases, calculation of renormalization functions (RFs) can be
carried out using lattice perturbation theory, which proves to be
extremely helpful in cases where there is a mixing with operators of
equal or lower dimension, such as the chromomagnetic operator
\cite{Constantinou:2014tea,Constantinou:2015ela} and the operator
measuring the glue of the nucleon~\cite{Alexandrou:2013tfa,Gluon}. 
However, perturbation theory is reliable for a limited range of values
of the coupling constant, $g$, and of the renormalization scale,
$\mu$. For this reason, a non-perturbative computation on the RFs is
preferable.

In this work we will combine both perturbative and non-perturbative
computations in order to obtain an improved evaluation for the RFs of the
quark field, ultra-local and one-derivative fermion operators within
the twisted mass formulation of Wilson lattice QCD~\cite{Frezzotti:2000nk}.
In particular, we compute lattice artifacts to all orders in the
lattice spacing, $a$, using one-loop perturbation theory and we
subtract them from the non-perturbative results for the RFs. This
subtraction suppresses lattice artifacts considerably depending on the
operator under study and leads to a more accurate determination of the
renormalization functions. We show that lattice artifacts are
non-negligible in  most cases, and are significantly larger than
statistical errors.

We use the Rome-Southampton method (RI$'$ scheme)~\cite{Martinelli:1994ty} 
to compute the renormalization coefficients of arbitrary quark-antiquark 
operators non-perturbatively. In this approach the renormalization
conditions are defined similarly in perturbative and
non-perturbative calculations. The RFs are obtained for different
values of the renormalization scale, and on several ensembles
corresponding to different pion masses, so that the chiral limit can
be safely taken. Since the goal is to make contact with
phenomenological and experimental studies, which almost exclusively
refer to operators renormalized in the ${\overline{\rm MS}}$ scheme,
one needs the renormalization functions leading from the bare operators
on the lattice to the ${\overline{\rm MS}}$ operators in the
continuum. The conversion to the ${\overline{\rm MS}}$ and the
evolution to a reference scale of 2~GeV is performed using 
three-loop perturbation theory.

The paper is organized as follows: Section \ref{sec2} presents the
lattice formulation and gives details on the gauge configurations and
the parameters of each ensemble. Section \ref{sec2}  includes the
definition of the operators under study, as well as the renormalization
conditions for the RI$'$ scheme. The methodolody of the non-perturbative 
computation is described in Section \ref{sec3}. Section \ref{sec4}
focuses on the perturbative procedure for the evaluation of the
one-loop lattice artifacts to all orders in the lattice spacing
denoted by ${\cal O}(g^2\,a^\infty)$. This is a crucial component of
this work, since the subtraction of the ${\cal O}(g^2\,a^\infty)$
contributions from the non-perturbative estimates of the RFs leads to
removal of the bulk of lattice artifacts. The main part of the paper
is Section \ref{sec5}, which presents the results of this work,
including their chiral extrapolation, the conversion to the
$\overline{\rm MS}$ scheme and the evolution to a reference scale of 2
GeV, via the intermediate Renormalization Group Invariant
scheme. Particular focus is given to the ${\cal O}(g^2\,a^\infty)$-corrected 
data for the RFs, and we show, for selected cases, a comparison with
the ${\cal O}(g^2\,a^2)$-corrected expressions. The final values of
the chirally extrapolated RFs at the limit $(a\,p)^2 \to 0$ are
presented in Table~\ref{tab3}. In Section~\ref{sec6} we give our
conclusions. For completeness we provide in Appendix \ref{appA} all
necessary formulae for the conversion to the $\overline{\rm MS}$
scheme, and in Appendix \ref{appB} we present the ${\cal
  O}(g^2\,a^\infty)$-corrected RFs for all the previous twisted mass
fermions ensembles~\cite{Alexandrou:2010me,Alexandrou:2012mt},
recomputed in the framework of this work.

\section{Formulation}
\label{sec2}

\subsection{Simulation details}

The gauge field configurations were generated by the European Twisted
Mass Collaboration (ETMC) employing the twisted mass fermion
action. We use ensembles generated with $N_f{=}4$ light degenerate
quarks~\cite{ETM:2011aa} with the  twisted mass action, as well as
$N_f{=}2$ degenerate quarks~\cite{Abdel-Rehim:2014nka}, in which a
clover term is also included in the fermion action with
$\csw{=}1.57551$. We note that the renormalization functions computed
using the $N_f{=}4$ ensembles will be applied to renormalize matrix
elements computed using $N_f{=}2{+}1{+}1$ gauge field
configurations~\cite{Alexandrou:2013joa}.
We adopt the RI$'$ renormalization scheme, which is mass independent,
and consequently the RFs are defined at zero quark mass. For this
reason, the $N_f{=}2{+}1{+}1$ ensembles cannot be used to compute the
RFs due to the fact that the mass of the strange and charm quarks are
fixed to their physical values, and extrapolation to the chiral limit
is not possible. Therefore, in order to compute the renormalization
functions needed to obtain physical observables, ETMC has generated
$N_f{=}4$ ensembles at the same values of $\beta$ so that the chiral
limit can be taken. Details on the simulations can be found in
Refs.~\cite{Abdel-Rehim:2013yaa,Abdel-Rehim:2014nka}.

Automatic ${\cal O}(a)$ improvement for twisted mass fermions may be
achieved with maximal twist, by tuning the $m_{\rm PCAC}$ quark mass
to zero. For the case of $N_f{=}4$ configurations, achieving maximal
twist is difficult particularly if the lattice spacing is not very
fine, which is associated with a change in the slope of $m_{\rm PCAC}$
with respect to $1/(2\kappa)$~~\cite{Dimopoulos:2011wz}. In order to
tackle this issue, Monte Carlo simulations are performed in pairs not
exactly at maximal twist but with opposite values of $m_{\rm
  PCAC}$. As proposed in Ref.~\cite{Frezzotti:2003ni}, by averaging
the RFs computed on ensembles with opposite values of $m_{\rm PCAC}$,
${\cal O}(a)$ improvement is achieved (see also
Ref.~\cite{Carrasco:2014cwa} and references therein). 

In the gluon sector we use, for all ensembles, the Iwasaki improved gauge
action~\cite{Weisz:1982zw}, which includes besides the plaquette term
$U^{1\times1}_{x,\mu,\nu}$ also rectangular $(1\times2)$ Wilson loops
$U^{1\times2}_{x,\mu,\nu}$
\be
    S_g =  \frac{\beta}{3}\sum_x\Biggl(  b_0\sum_{\substack{
      \mu,\nu=1\\\mu<\nu}}^4\left \{1-\re\tr(U^{1\times 1}_{x,\mu,\nu})\right \}\Bigr. 
     \Bigl.\,+
    \,b_1\sum_{\substack{\mu,\nu=1\\\mu\neq\nu}}^4\left \{1
    -\re\tr(U^{1\times2}_{x,\mu,\nu})\right \}\Biggr)\,  
\label{Symanzik}
\ee
with $\beta{=}2\,N_c/g_0^2$, $b_1=-0.331$ and the
(proper) normalization condition $b_0=1-8b_1=3.648$. 

The simulation details, the parameters and the values of the pion
mass~\cite{pion_masses} of each ensemble used in this work are given
in Tables \ref{tab1} - \ref{tab2}, for the $N_f{=}4$ and $N_f{=2}$
ensembles, respectively. The values of the lattice spacing have been
determined using the nucleon mass \cite{Alexandrou:2013joa, Abdel-Rehim:2015jna}.

\begin{table}[!h]
\begin{center}
\begin{tabular}{ccccc}
\hline
\hline
$a \mu$ & $\kappa$ & $a \mu^{sea}_{\rm PCAC}$ & $a M_{PS}$ & lattice size\\
\hline
\hline
$\,\,\,$        $\,\,\,$   &             & $\beta=1.90$, $a=0.0934$ fm  &        &           \\ \hline
$\,\,\,$  0.0080$\,\,\,$   &  0.162689   & $+$0.0275(4)     & 0.280(1)   & $24^3 \times 48$ \\  %A4p
$\,\,\,$        $\,\,\,$   &  0.163476   & $-$0.0273(2)     & 0.227(1)   \\ \hline              %A4m
$\,\,\,$  0.0080$\,\,\,$   &  0.162876   & $+$0.0398(1)     & 0.279(2)   & $24^3 \times 48$ \\  %A1p
$\,\,\,$        $\,\,\,$   &  0.163206   & $-$0.0390(1)     & 0.241(1)   \\ \hline              %A1m
$\,\,\,$        $\,\,\,$   &             & $\beta=1.95$, $a=0.082$ fm  &         &          \\ \hline
$\,\,\,$  0.0020$\,\,\,$   &  0.160524   & $+$0.0363(1)     &    & $24^3 \times 48$ \\          %B8p
$\,\,\,$        $\,\,\,$   &  0.161585   & $-$0.0363(1)     &    \\ \hline                      %B8m
$\,\,\,$  0.0085$\,\,\,$   &  0.160826   & $+$0.0191(2)     & 0.277(2)   & $24^3 \times 48$ \\  %B2p
$\,\,\,$        $\,\,\,$   &  0.161229   & $-$0.0209(2)     & 0.259(1)   \\ \hline              %B2m
$\,\,\,$  0.0180$\,\,\,$   &  0.160826   & $+$0.0163(2)     & 0.317(1)   & $24^3 \times 48$ \\  %B3p
$\,\,\,$        $\,\,\,$   &  0.161229   & $-$0.0160(2)     & 0.292(1)   \\ \hline              %B3m
$\,\,\,$        $\,\,\,$   &             & $\beta=2.10$, $a=0.064$ fm    &      &           \\ \hline
$\,\,\,$  0.0030$\,\,\,$   &  0.156042   & $+$0.0042(1)     & 0.127(2)   & $32^3 \times 64$ \\  %C2p
$\,\,\,$        $\,\,\,$   &  0.156157   & $-$0.0040(1)     & 0.129(3)   \\ \hline              %C2m
$\,\,\,$  0.0046$\,\,\,$   &  0.156017   & $+$0.0056(1)     & 0.150(2)   & $32^3 \times 64$ \\  %C3p
$\,\,\,$        $\,\,\,$   &  0.156209   & $-$0.0059(1)     & 0.160(4)   \\ \hline              %C3m
$\,\,\,$  0.0064$\,\,\,$   &  0.155983   & $+$0.0069(1)     & 0.171(1)   & $32^3 \times 64$ \\  %C4p
$\,\,\,$        $\,\,\,$   &  0.156250   & $-$0.0068(1)     & 0.180(4)   \\ \hline              %C4m
$\,\,\,$  0.0078$\,\,\,$   &  0.155949   & $+$0.0082(1)     & 0.188(1)   & $32^3 \times 64$ \\  %C5p
$\,\,\,$        $\,\,\,$   &  0.156291   & $-$0.0082(1)     & 0.191(3) \\                       %C5m
\hline
\hline
\end{tabular}
\caption{Simulation details for the $N_f{=}4$ ensembles of twisted
mass fermions. The lattice spacing is determined using the nucleon mass
computed with $N_f{=}2{+}1{+}1$ twisted mass configurations at the same
values of $\beta$.} 
\label{tab1}
\end{center}
\end{table}

\begin{table}[!h]
\begin{center}
\begin{tabular}{cccc}
\hline
\hline
$a \mu$ & $\kappa$  & $a M_{PS}$ & lattice size\\
\hline
\hline
\multicolumn{4}{c}{$\beta=2.10$,  $a=0.093$ fm,  $\csw=1.57551$ }\\
\hline
0.0009   &  0.13729    & 0.0621(2)  & $48^3 \times 96$ \\  
0.0030   &  0.1373     & 0.110(4)   & $24^3 \times 48$ \\  
0.0060   &  0.1373     & 0.160(4)   & $24^3 \times 48$ \\  
\hline
\hline
\end{tabular}
\caption{Simulation details for the $N_f{=}2$ twisted mass ensembles
  with a clover term.  The lattice spacing is determined using the nuleon mass
computed using the same $N_f{=}2$ ensembles.}
\label{tab2}
\end{center}
\end{table}

The number of configurations in each ensemble varies between 10 to
50 separated by 20-100 trajectories, depending on the ensemble.
The small size of these ensembles, is more than sufficient for use of
the momentum source method, which offers high statistical accuracy,
easily below 0.5$\%$ even for 10 configurations (see Section~\ref{sec3}). 
In our computation, we mostly use ``democratic momenta'' in the spatial
direction, such as:
\be
(a\,p) \equiv 2\pi \left(\frac{n_t}{L_t}+\frac{1}{2\,L_t},
\frac{n_x}{L_s},\frac{n_x}{L_s},\frac{n_x}{L_s}\right)\,,
\ee 
where $L_t$ ($L_s$) is the temporal (spatial) extent of the lattice
and $n_t$ and $n_x$ take values within the range: 
\be
n_t \,\epsilon\, [2, 20]\,,\quad n_x\,\epsilon\, [1, 10]\,,
\ee
depending on the lattice size of each ensemble, so that they
correspond to momentum up to $(a\,p)^2{\sim} 7$. To fill in some gaps
between the momentum ranges we also include a few non-democratic momenta
of the form $(n_t,n_x,n_x,n_x\pm 1)$, which show similar behaviour with
neighbouring democratic momenta. 

%%%%%%%%%%%%%%%%%%%%%%%%%%%%%%%%%%%%%%%%%%%%%%%%%%%%%%%%%%%%%

\subsection{Definition of operators and renormalization prescription}

In this work we consider ultra-local fermion operators:
\begin{eqnarray}
   \Op_S^a &= \bar \chi \tau^a \chi           &= \begin{cases} \bar \psi \tau^a          \psi   & a=1,2 \\
                                                             -i\bar \psi \gamma_5 \eins  \psi   & a=3 \end{cases} \\
   \Op_P^a &= \bar \chi \gamma_5\tau^a \chi   &= \begin{cases} \bar \psi \gamma_5 \tau^a \psi   & a=1,2 \\
                                                             -i\bar \psi          \eins  \psi   & a=3 \end{cases} \\
   \Op_V^a &= \bar \chi \gamma_\mu\tau^a \chi &= \begin{cases} \bar \psi  \gamma_5\gamma_\mu \tau^2 \psi   & a=1 \\
                                                              -\bar \psi  \gamma_5\gamma_\mu \tau^1 \psi   & a=2 \\
                                                               \bar \psi  \gamma_\mu         \tau^3 \psi   & a=3 \end{cases} \\
   \Op_A^a &= \bar \chi \gamma_5\gamma_\mu\tau^a \chi &= \begin{cases} \bar \psi  \gamma_\mu        \tau^2 \psi   & a=1 \\
                                                                      -\bar \psi  \gamma_\mu        \tau^1 \psi   & a=2 \\
                                                                       \bar \psi  \gamma_5\gamma_\mu\tau^3 \psi   & a=3 \end{cases}\\
   \Op_T^a &= \bar \chi \smn\tau^a \chi &= \begin{cases} \bar \psi  \smn      \tau^a \psi   & a=1,2 \\
                                                                  -i\bar \psi  \gamma_5\smn\eins\psi   & a=3 \end{cases}\\
   \Op_{Tp}^a &= \bar \chi \gamma_5\smn\tau^a \chi &= \begin{cases} \bar \psi \gamma_5 \smn      \tau^a \psi   & a=1,2 \\
                                                                  -i\bar \psi \smn\eins\psi   & a=3 \end{cases}
\end{eqnarray}
and the following one-derivative fermion operators:
\begin{eqnarray}
\Op_{\rm DV}^{\{\mu\,\nu\}} &= \overline \chi \gamma_{\{\mu}\overleftrightarrow D_{\nu\}}\tau^a \chi 
                                              &= \begin{cases} \overline \psi  \gamma_5\gamma_{\{\mu}\overleftrightarrow D_{\nu\}} \tau^2 \psi   & a=1 \\
                                                              -\overline \psi  \gamma_5\gamma_{\{\mu}\overleftrightarrow D_{\nu\}} \tau^1 \psi   & a=2 \\
                                                               \overline \psi  \gamma_{\{\mu}\overleftrightarrow D_{\nu\}}         \tau^3 \psi   & a=3 \end{cases} \\[3ex]
\Op_{\rm DA}^{\{\mu\,\nu\}} &= \overline \chi \gamma_5\gamma_{\{\mu}\overleftrightarrow D_{\nu\}}\tau^a \chi 
                                              &= \begin{cases} \overline \psi  \gamma_{\{\mu}\overleftrightarrow D_{\nu\}} \tau^2 \psi   &\quad a=1 \\
                                                              -\overline \psi  \gamma_{\{\mu}\overleftrightarrow D_{\nu\}} \tau^1 \psi   & \quad a=2 \\
                                                               \overline \psi  \gamma_5\gamma_{\{\mu}\overleftrightarrow D_{\nu\}} \tau^3 \psi   & \quad a=3 \end{cases}\\[3ex]
\Op_{\rm DT}^{\mu\,\{\nu\,\rho\}} &= \overline \chi \gamma_5\sigma_{\mu\{\nu}\overleftrightarrow D_{\rho\}}\tau^a \chi 
                                              &= \begin{cases} \overline \psi  \gamma_5\sigma_{\mu\{\nu}\overleftrightarrow D_{\rho\}}\tau^a \psi   & a=1,2 \\
                         -i\,\overline \psi  \sigma_{\mu\{\nu}\overleftrightarrow D_{\rho\}}\eins \psi            & a=3 \end{cases}\,,
\end{eqnarray}
all given in the twisted and physical basis as shown above. The
covariant derivative is defined as:
\be
\Dlr = \frac{1}{2}\Big[\frac{(\overrightarrow\nabla_{\mu} +
    \overrightarrow\nabla_{\mu}^{*})}{2} -  \frac{(\overleftarrow\nabla_{\mu} +
    \overleftarrow\nabla_{\mu}^{*})}{2} \Big]\,
\ee
where
\be
\overrightarrow\nabla_\mu \psi(x)= \frac{1}{a}\biggl[U_\mu(x)\psi(x+a\hat{\mu})-\psi(x)\biggr]
\hspace*{0.5cm} {\rm and}\hspace*{0.5cm} 
\overrightarrow\nabla^*_{\mu}\psi(x)=-\frac{1}{a}\biggl[U^\dagger_{\mu}(x-a\hat{\mu})\psi(x-a\hat{\mu})-\psi(x)\biggr]
\quad 
\ee
and 
\be
\overline\psi(x)\overleftarrow\nabla_\mu = \frac{1}{a}\biggl[\overline\psi(x+a\hat{\mu})U^\dagger_\mu(x)-\overline\psi(x)\biggr]
\hspace*{0.5cm} {\rm and}\hspace*{0.5cm} 
\overline\psi(x)\overleftarrow\nabla^*_{\mu}=-\frac{1}{a}\biggl[\overline\psi(x-a\hat{\mu})U_{\mu}(x-a\hat{\mu})-\overline\psi(x)\biggr]
\quad .
\ee
For completeness we include in the list  $\Op_{Tp}^a$ even though its
components are related to those of $\Op_{T}^a$. 
We denote the corresponding RFs of the ultra-local fermion bilinears by 
$\ZS^a,\,\ZP^a,\,\ZV^a,\,\ZA^a,\ZT^a,Z_{\rm Tp}^a$.
In a massless renormalization scheme, such as the RI$'$, the RFs are
defined in the chiral limit, where iso-spin symmetry is
recovered. Hence, the renormalization functions become independent of
the isospin index $a=1,2,3$ and we drop the $a$ index from here
on. Still note that, for instance, the physical $\bar\psi \gamma_\mu
\tau^1 \psi$ is renormalized with $\ZA$ while $\bar\psi \gamma_\mu
\tau^3 \psi$ needs $\ZV$, which differ from each other even in the
chiral limit.

The one-derivative operators are symmetrized over two Lorentz indices
and are made traceless:
\be
\Op^{\{\sigma\,\tau\}} \equiv \frac{1}{2}\Big(\Op^{\sigma\,\tau}+\Op^{\tau\,\sigma}
\Big) - \frac{1}{4}\delta^{\sigma\,\tau} \sum_\lambda \Op^{\lambda\,\lambda}\,,
\ee
which avoids mixing with lower dimension operators. The corresponding
RFs of the one-derivative operators are denoted by $Z_{\rm DV}^a$,
$Z_{\rm DA}^a$, $Z_{\rm DT}^a\,$. The one-derivative operators fall
into different irreducible representations of the hypercubic group,
depending on the choice of indices. Hence, we distinguish among them 
according to the following
\begin{eqnarray}
   \Op_{\rm DV1} &=& \Op_{\rm DV} \ {\rm with} \ \mu=\nu \\
\Op_{\rm DV2} &=& \Op_{\rm DV} \ {\rm with} \ \mu\neq\nu \\
   \Op_{\rm DA1} &=& \Op_{\rm DA} \ {\rm with} \ \mu=\nu \\
   \Op_{\rm DA2} &=& \Op_{\rm DA} \ {\rm with} \ \mu\neq\nu\\
   \Op_{\rm DT1} &=& \Op_{\rm DT} \ {\rm with} \ \mu\neq\nu=\rho\\
   \Op_{\rm DT2} &=& \Op_{\rm DT} \ {\rm with} \ \mu\neq\nu\neq\rho\neq\mu\,.
\end{eqnarray}
Thus, $Z_{\rm DV1}$ will be different from $Z_{\rm DV2}$, but
renormalized matrix elements of the two corresponding operators will
be components of the same tensor in the continuum limit. 

\vspace{0.75cm}

The renormalization functions are computed in the RI$'$ scheme at
different renormalization scales, $\mu$. The RFs are determined by
imposing the following conditions:
\begin{eqnarray}
   \Zq = \frac{1}{12} {\rm Tr} \left[(S^L(p))^{-1}\, S^{{\rm Born}}(p)\right] \Bigr|_{p^2=\mu^2}  \label{Zq_cond}\\[2ex]
   \Zq^{-1}\,Z_{\cal O}\,\frac{1}{12} {\rm Tr} \left[\Gamma^L(p)
     \,\Gamma^{{\rm Born}-1}(p)\right] \Bigr|_{p^2=\mu^2} &=& 1\, ,
\label{renormalization cond}
\end{eqnarray}
where the momentum $p$ is set to the renormalization scale $\mu$. The
trace is taken over spin and color indices, $S^{{\rm Born}}$ is the
tree-level result for the propagator, and $\Gamma^{{\rm Born}}$ is the
tree-level expressions for the fermion operators, that is 
\be
\Gamma^{{\rm Born}}(p) = \openone,\,\,\gamma_5,\,\, \gamma_\mu,\,\,\gamma_5\,\gamma_\mu,
\,\, \gamma_5\,\smn,\,\,\smn
\label{meth2}
\ee
for the ultra-local bilinears, and
\bea
{\cal O}^{\{\mu\nu\}}_{\rm DV} &=& \frac{1}{2}\Big[\overline\Psi\,\gamma_{\mu}\,\Dlr_{\nu}\,\Psi + 
\overline\Psi\,\gamma_{\nu}\,\Dlr_{\mu}\,\Psi \Big] 
-\frac{1}{4} \delta_{\mu\nu} \sum_\tau \overline\Psi\,\gamma_\tau\,\Dlr_{\tau}\,\Psi  \label{extV}\\ [2ex]
{\cal O}^{\{\mu\nu\}}_{\rm DA} &=& \frac{1}{2}\Big[\overline\Psi\,\gamma_5\gamma_{\mu}\,\Dlr_{\nu}\,\Psi + 
\overline\Psi\,\gamma_5\gamma_{\nu}\,\Dlr_{\mu}\,\Psi \Big] 
-\frac{1}{4} \delta_{\mu\nu} \sum_\tau \overline\Psi\,\gamma_5\gamma_\tau\,\Dlr_{\tau}\,\Psi \label{extA}\\ [2ex]
{\cal O}^{\mu\{\nu\rho\}}_{\rm DT} &=& \frac{1}{2}\Big[\overline\Psi\,\gamma_5\sigma_{\mu\nu}\,\Dlr_{\rho}\,\Psi + 
\overline\Psi\,\gamma_5\sigma_{\mu\rho}\,\Dlr_{\nu}\,\Psi \Big] 
-\frac{1}{4} \delta_{\nu\rho} \sum_\tau \overline\Psi\,\gamma_5\sigma_{\mu\tau}\,\Dlr_{\tau}\,\Psi 
\label{extT}
\eea
for the one-derivative operators. The presence of $S^{{\rm Born}}$ and
$\Gamma^{{\rm Born}}$ ensure that $\Zq = 1$, $Z_{\cal O} = 1$ when the
gauge field is set to unity. The RI$'$ values for the RFs are
translated to the ${\overline{\rm MS}}$ scheme at $\mu=$2~GeV using an
intermediate Renormalization Group Invariant scheme.

\section{Non-perturbative calculation}
\label{sec3}

For the non-perturbative evaluation we follow the same
procedure as our previous work~\cite{Alexandrou:2010me,Alexandrou:2012mt}, 
and here we summarize the important steps of the calculation.  We
first write the operators  in the form 
\begin{equation}
   \Op(z) = \sum_{z'} \overline u(z) \J(z,z') d(z')\, ,
\end{equation}
where $u$ and $d$ denote quark fields in the physical basis and $\J$
denotes the operator we are interested in. For example $\J(z,z') =
\delta_{z,z'} \gamma_\mu$ corresponds to the local vector current. 
For each operator we define a bare vertex function given by
\begin{equation}\label{vfun}  
   G(p) = \frac{a^{12}}{V}\sum_{x,y,z,z'} e^{-ip(x-y)} \langle u(x) \overline u(z) \J(z,z') d(z') \overline d(y) \rangle \, ,
\end{equation}
where $p$ is a momentum allowed by the boundary conditions, $V$ is the
lattice volume, and the gauge average, denoted by the brackets, is
performed over gauge-fixed configurations. The Dirac and color indices
of $G(p)$ are suppressed for simplicity.

We employ the approach, introduced in Ref.~\cite{Gockeler:1998ye},
which uses directly Eq.~(\ref{vfun}) without employing translation
invariance
\footnote{In an alternative approach that relies on translation
invariance, one may shift the coordinates of the correlators in
Eq.~(\ref{vfun}) to position $z=0$~\cite{Carrasco:2014cwa}.},
and one must now use a source that is momentum dependent but
can couple to any operator. For twisted mass fermions, we use the
symmetry $S^u(x,y)=\gamma_5S^{d\dagger}(y,x)\gamma_5$ between the $u-$
and $d-$quark propagators, and therefore, with a single inversion one
can extract the vertex function for a single momentum.
The advantage of the momentum source approach is a high statistical
accuracy and the evaluation of the vertex for any operator at no
significant additional computational cost. The drawback is that 
we need a new inversion for each momentum. We fix to Landau gauge
using a stochastic over-relaxation algorithm~\cite{deForcrand:1989im},
converging to a gauge transformation which minimizes the functional
\begin{equation}
   F = \sum_{x,\mu} {\rm Re}\ {\rm tr} \left[ U_\mu(x) + U^\dagger_\mu(x-\hat\mu)\right] \, .
\end{equation}
The propagator in momentum space, in the physical basis, is defined by
\begin{equation}\label{pprop}
   S^u(p) = \frac{a^8}{V}\sum_{x,y} e^{-ip(x-y)} \left\langle u(x) \overline u(y) \right\rangle\, , \qquad
   S^d(p) = \frac{a^8}{V}\sum_{x,y} e^{-ip(x-y)} \left\langle d(x) \overline d(y) \right\rangle \, ,
\end{equation}
and an amputated vertex function is given by
\begin{equation}
   \Gamma(p) = (S^u(p))^{-1}\, G(p)\, (S^d(p))^{-1} \, .
\label{vertexfunction}
\end{equation}
Finally, the corresponding renormalized quantities are assigned the values
\begin{equation}
   S_R(p)      = \Zq S(p) \, , \qquad \qquad
   \Gamma_R(p) = \Zq^{-1} Z_\Op \Gamma(p) \quad.
\end{equation}
In the twisted basis at maximal twist, Eq.~(\ref{vfun}) takes the form
\begin{equation}\label{vfun_tm}
   G(p) = \frac{a^{12}}{4V}\sum_{x,y,z,z'} e^{-ip(x-y)} \left\langle(\eins+i\gamma_5) u(x) \overline u(z)(\eins+i\gamma_5) \J(z,z') (\eins-i\gamma_5) d(z')\
 \overline d(y)(\eins-i\gamma_5) \right\rangle \, ,
\end{equation}
which simplifies when using the relation between the u- and d-quark
propagators, that is $\s^u(x,z)=\gamma_5 {\s^d}^\dagger(z,x)\gamma_5$.
After integration over the fermion fields it becomes
\begin{equation}
   G(p) = -\frac{a^{12}}{4V} \sum_{z,\,z'} \left\langle (\eins-i\gamma_5){\breve {\s^d}}^\dagger(z,p)(\eins-i\gamma_5) \J(z,z')
   (\eins-i\gamma_5)\breve \s^d(z',p)(\eins-i\gamma_5) \right\rangle^G \, ,
\end{equation}
where $\langle ... \rangle^G$ denotes the integration over gluon fields,
and $\breve \s(z,p) = \sum_y e^{ipy} \s(z,y)$ is the Fourier
transformed propagator with respect to one of its arguments, on a
particular gauge background. It can be obtained by inversion using the
Fourier source  
\begin{equation}
   b_\alpha^a(x) = e^{ipx} \delta_{\alpha \beta}\delta_{a b} \, ,
\end{equation}
for all Dirac $\alpha$ and color $a$ indices. The propagators in the
physical basis given in Eq.~(\ref{pprop}) can be obtained from
\begin{eqnarray}\label{pprop_tm}
   S^d(p) &=& \phantom{-}\frac{1}{4} \sum_z e^{-ipz} \langle (\eins-i\gamma_5)\breve \s^d(z,p)(\eins-i\gamma_5) \rangle^G \nonumber \\
   S^u(p) &=&           -\frac{1}{4} \sum_z e^{+ipz} \langle (\eins-i\gamma_5){\breve {\s^d}}^\dagger(z,p)(\eins-i\gamma_5) \rangle^G \, ,
\end{eqnarray}
which only need 12 inversions (instead of 24) despite the occurrence
of both $u$ and $d$ quarks in the original expression. We evaluate
Eq.~(\ref{vfun_tm}) and Eq.~(\ref{pprop_tm}) for each momentum
separately employing Fourier sources over a range of $(a\,p)^2$ for
which perturbative results can be trusted and finite $a$ corrections
are reasonably small. The amputated vertex functions of
Eq.~(\ref{vertexfunction}) computed for each operator, as well as the
inverse quark propagator, enter the renormalization prescription of
Eqs.~(\ref{Zq_cond}) - (\ref{renormalization cond}).

\section{One-loop calculation of artifacts to all orders in the
  lattice spacing}
\label{sec4}

An improvement over previous work~\cite{Alexandrou:2010me,Alexandrou:2012mt}, 
where we
evaluated the ${\cal O}(g^2\,a^2)$ perturbative artifacts, is the computation of the  one-loop perturbative
artifacts to all orders in the lattice spacing, ${\cal O}(g^2\,a^\infty)$.
These artifacts 
 are unavoidably present in the non-perturbative vertex functions
of the fermion propagator and fermion operators under study. In our
previous work \cite{Alexandrou:2010me,Alexandrou:2012mt}, the 
${\cal O}(g^2\,a^2)$ perturbative artifacts were subtracted from the
non-perturbative RFs, leading to improved estimates. However, for
large values of the scale $(a\,p)^2$, the ${\cal O}(g^2\,a^2)$ terms
tend to increase becoming, thus, unreliable.

As will be demonstrated by our results, the lattice
artifacts depend on the operator under study, as well as on several
parameters such as the coupling constant, the fermion and gluon
action parameters, the lattice size, the lattice spacing and the
renormalization scale. Thus, a proper subtraction of the finite
lattice size effects from the non-perturbative values requires a
separate perturbative evaluation of the ${\cal O}(g^2\,a^\infty)$
terms for each ensemble and each value of the four-momentum,
in order to match our non-perturbative computation. In other words,
the ${\cal O}(g^2\,a^\infty)$ contributions that are subtracted from
each black circle point shown in the plots of Section~\ref{sec5}
requires a separate perturbative computation. Unlike the case of the
${\cal O}(g^2\,a^2)$-subtraction used in our previous work for the
renormalization functions~\cite{Constantinou:2009tr,Alexandrou:2010me,Alexandrou:2012mt},
the ${\cal O}(g^2\,a^\infty)$ contributions cannot be given in a
closed form. 

\begin{figure}[h]
\centerline{\psfig{figure=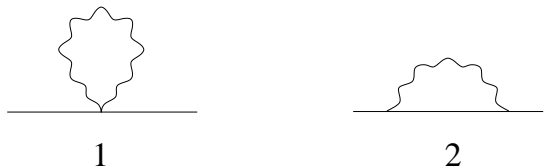,height=2.15truecm}}
\caption{One-loop diagrams contributing to the fermion
propagator. Wavy (solid) lines represent gluons (fermions).}
\label{fig1}
\end{figure}

\begin{figure}[h]
\centerline{\psfig{figure=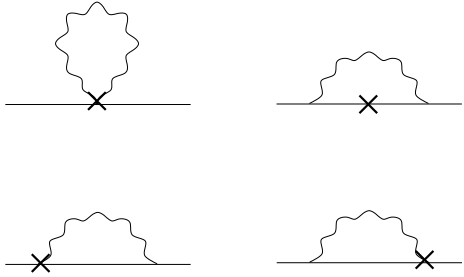,height=3.65truecm}}
\caption{One-loop diagrams contributing to the computation of the
fermion operators. A wavy (solid) line represents gluons
(fermions). A cross denotes an insertion of the operator under
study. In the case of the ultra-local operators, only the upper right
diagram contributes due to the absense of gluons in the vertices.}
\label{fig2}
\end{figure}

\bigskip
There are six Feynman diagrams that enter the perturbative computation:
Two for the fermion propagator and four for the fermion operators, as shown
in Figs.~\ref{fig1} - \ref{fig2}. The operator insersion is
represented by a cross. In this work we restrict ourselves to forward
matrix elements (i.e. 2-point Green's functions, zero momentum
operator insertions). The Feynman diagrams are evaluated using our
symbolic package in Mathematica, and details on the algebraic
operations can be found in Ref.~\cite{Constantinou:2009tr}.

As a general strategy in our perturbative computations for the RFs
we employ a variety of fermionic and gluonic actions, in order to
obtain results that are applicable to simulations performed by various 
research groups. In this paper, however, we only present the results for the
twisted mass action including a clover term. The latter is kept as a
free parameter and can be sent to zero as required for the $N_f{=}4$
ensembles presented in Table~\ref{tab1}. Although the clover parameter
is treated as free throughout the perturbative computation, for our
final estimates in the $N_f{=}2$ ensembles it is set to its
tree-level value suggested by one-loop perturbation theory, 
$c_{\rm sw} = 1$.

The computation of the ${\cal O}(g^2\,a^\infty)$ terms was first
employed by the QCDSF Collaboration~\cite{Gockeler:2010yr,Constantinou:2013ada}
for Clover fermions and Wilson gluons, and was later generalized to
include more complicated fermion and gluon actions
\cite{Constantinou:2014fka}. The main difference between the
computation of the ${\cal O}(g^2\,a^\infty)$ and the 
${\cal O}(g^2\,a^2)$ terms, is that the latter are extracted by
performing a Taylor expansion with respect to $a$. The 
${\cal O}(g^2\,a^\infty)$ terms, however, cannot be given in a closed
form in terms of $a$ (since it is included in the propagators) and a
separate calculation is performed for each value of the 
momentum, $(a\,p)^2$. Of course, from the resulting expression one
must omit the ${\cal O}(g^2\,a^0)$ terms since we are interested only
in the lattice artifacts. Note that, in most cases, the latter
contributions include logarithms (${\cal O}(g^2 \log(a))\,$), which are
also subtracted. In a nutshell, the lattice artifacts to all orders in
the lattice spacing are computed in the procedure summarized in the
following expressions, in which the ${\cal O}(g^2\,a^0)$ terms
computed in Ref.~\cite{Constantinou:2009tr}, are omitted: 
\bea
D{\cal Z}_q(a,p) = \left({\cal V}_q(a,p) - {\cal V}_q(0,p) \right) \Bigr|_{p^2=\mu^2} \\[2ex]
D{\cal Z}_{\cal O}(a,p) = \left(\frac{{\cal V}_q(a,p)}{{\cal V}_{\cal O}(a,p)}
                              - \frac{{\cal V}_q(0,p)}{{\cal V}_{\cal O}(0,p)} \right) \Bigr|_{p^2=\mu^2}\,,
\eea
where 
\be
{\cal V}_q(a,p) = \frac{1}{12} {\rm Tr} \left[(S^L(a,p))^{-1}\, S^{{\rm Born}}(p)\right]\,, \qquad
{\cal V}_{\cal O}(a,p) = \frac{1}{12} {\rm Tr} \left[\Gamma^L(a,p)\,\Gamma^{{\rm Born}-1}(p)\right]
\ee
and $S^L(a,p)$, $\Gamma^L(a,p)$ are the results up to one loop and to
all orders in $a$. Finally,  the perturbative ${\cal O}(g^2a^\infty)$-terms
are subtracted form the non-perturbative values
\bea
Z_q^{{\rm RI}',{\rm sub}}(p,a) = Z_q^{{\rm RI}'}(p,a) - D{\cal Z}_q(a,p) \\[2ex]
Z_{\cal O}^{{\rm RI}',{\rm sub}}(p,a) = Z_{\cal O}^{{\rm RI}'}(p,a) - D{\cal Z}_{\cal O}(a,p)\,.
\eea

\begin{figure}[!h]
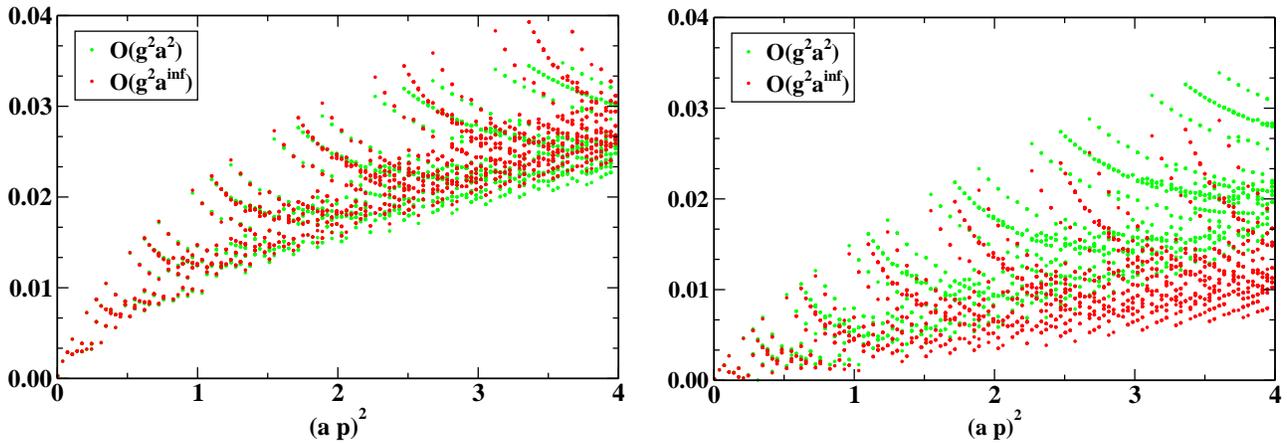

{\includegraphics[scale=0.345]{./Zq_csw_0.eps}}\,\,$\quad$
{\includegraphics[scale=0.345]{./Zq_csw_1.eps}}
\caption{$D{\cal Z}_q(a,p)/g^2$ as a function of $(a\,p)^2$ with the
Iwasaki gluon action and a $24^3\times48$ lattice for $c_{\rm sw}=0$
(left) and $c_{\rm sw}=1$ (right).}
\label{Zq_pert}
\end{figure}

\begin{figure}[!h]
{\includegraphics[scale=0.345]{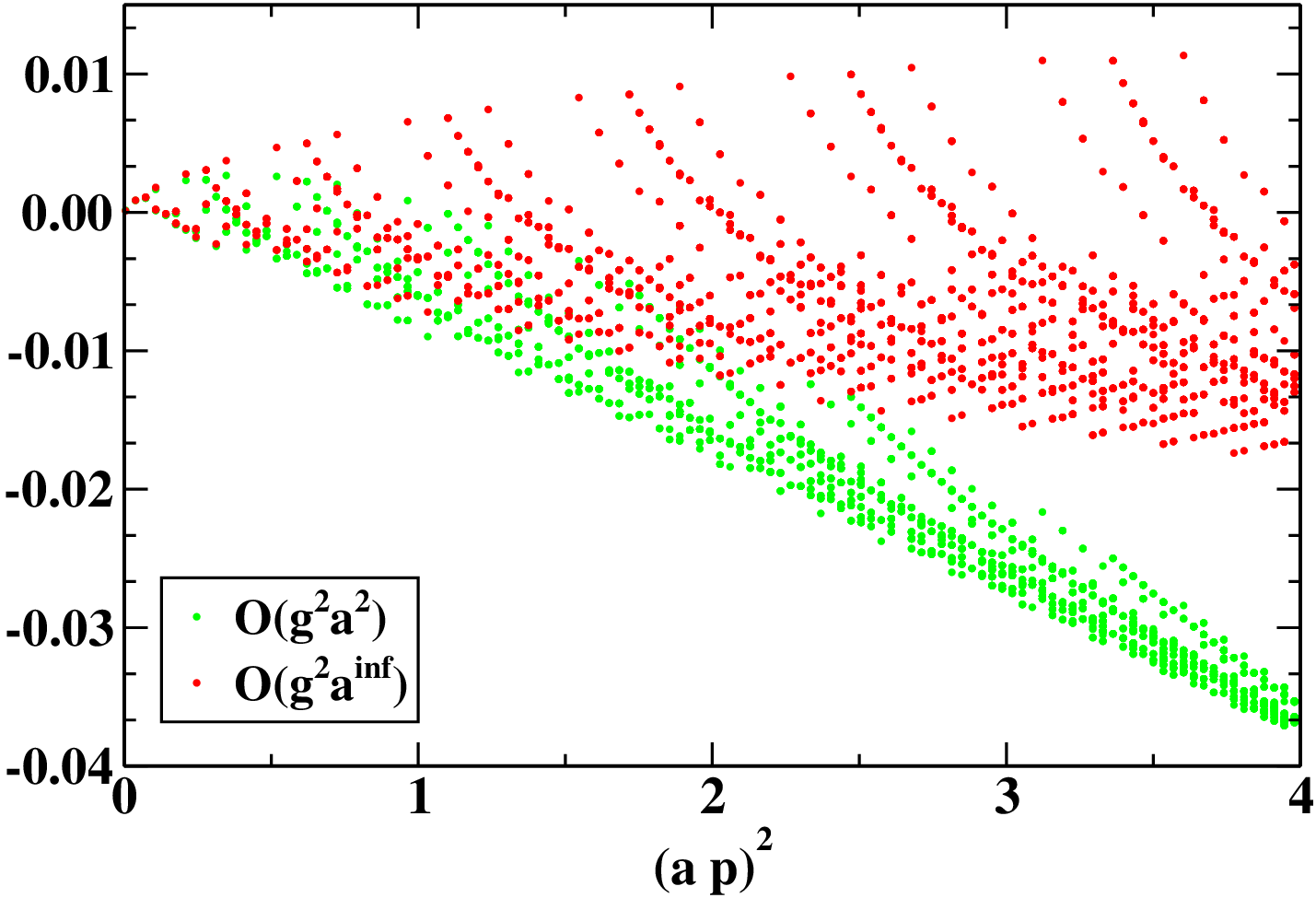}}$\quad$
{\includegraphics[scale=0.345]{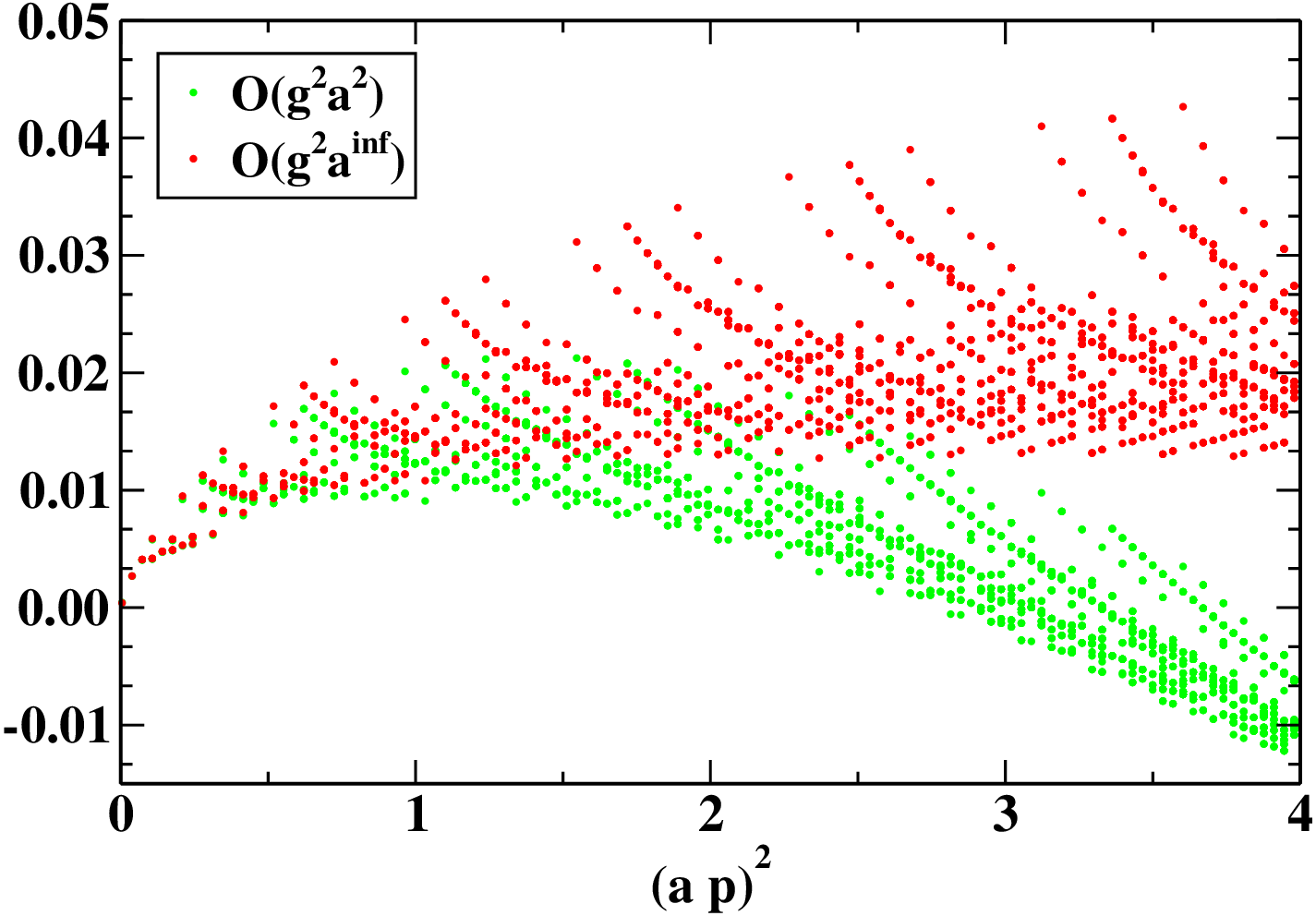}}
\caption{$D{\cal Z}_S(a,p)/g^2$ as a function of $(a\,p)^2$. The notation is the same as for Fig.~\ref{Zq_pert}.}
\label{ZS_pert}
\end{figure}

\begin{figure}[!h]
{\includegraphics[scale=0.345]{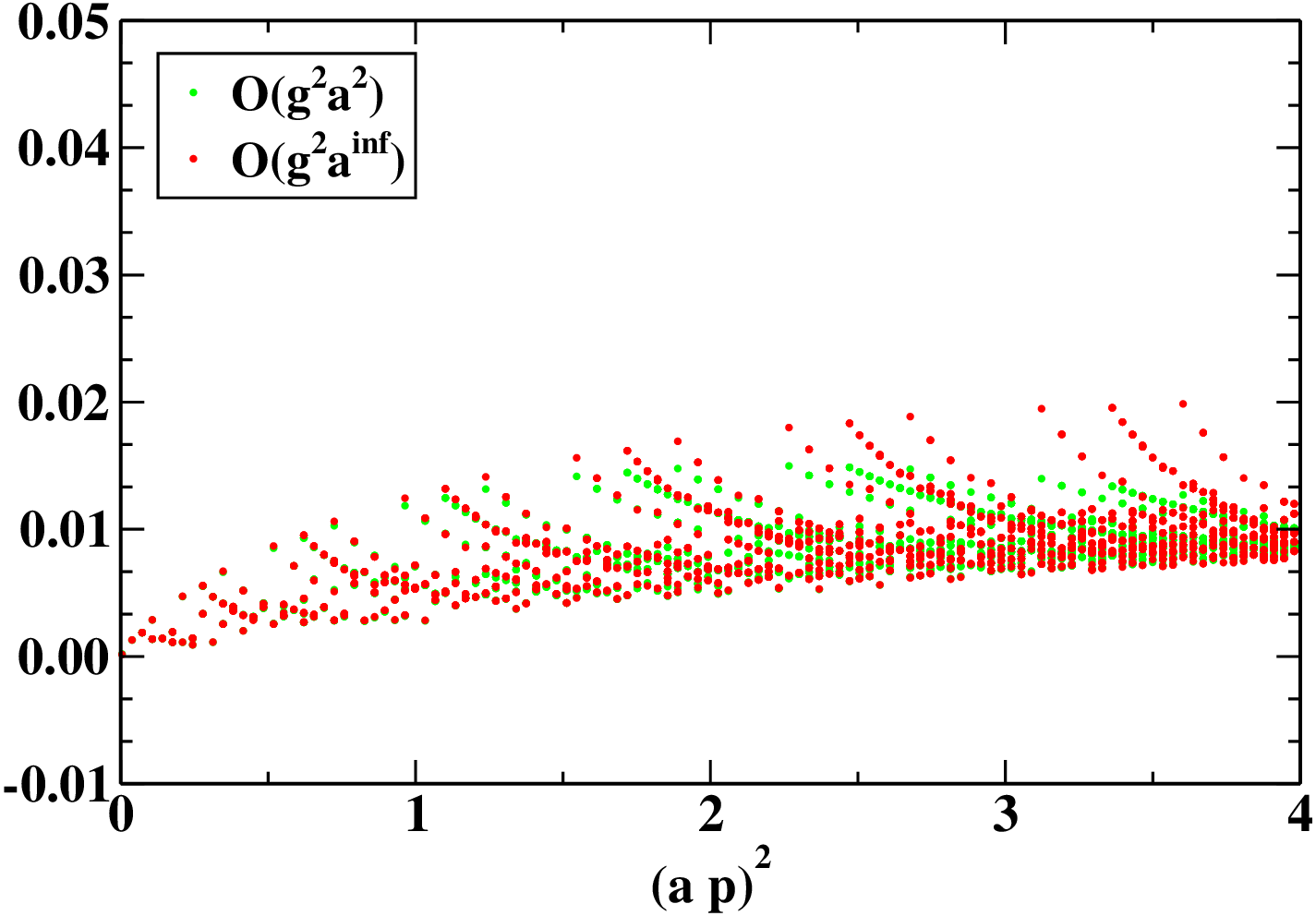}}$\quad$
{\includegraphics[scale=0.345]{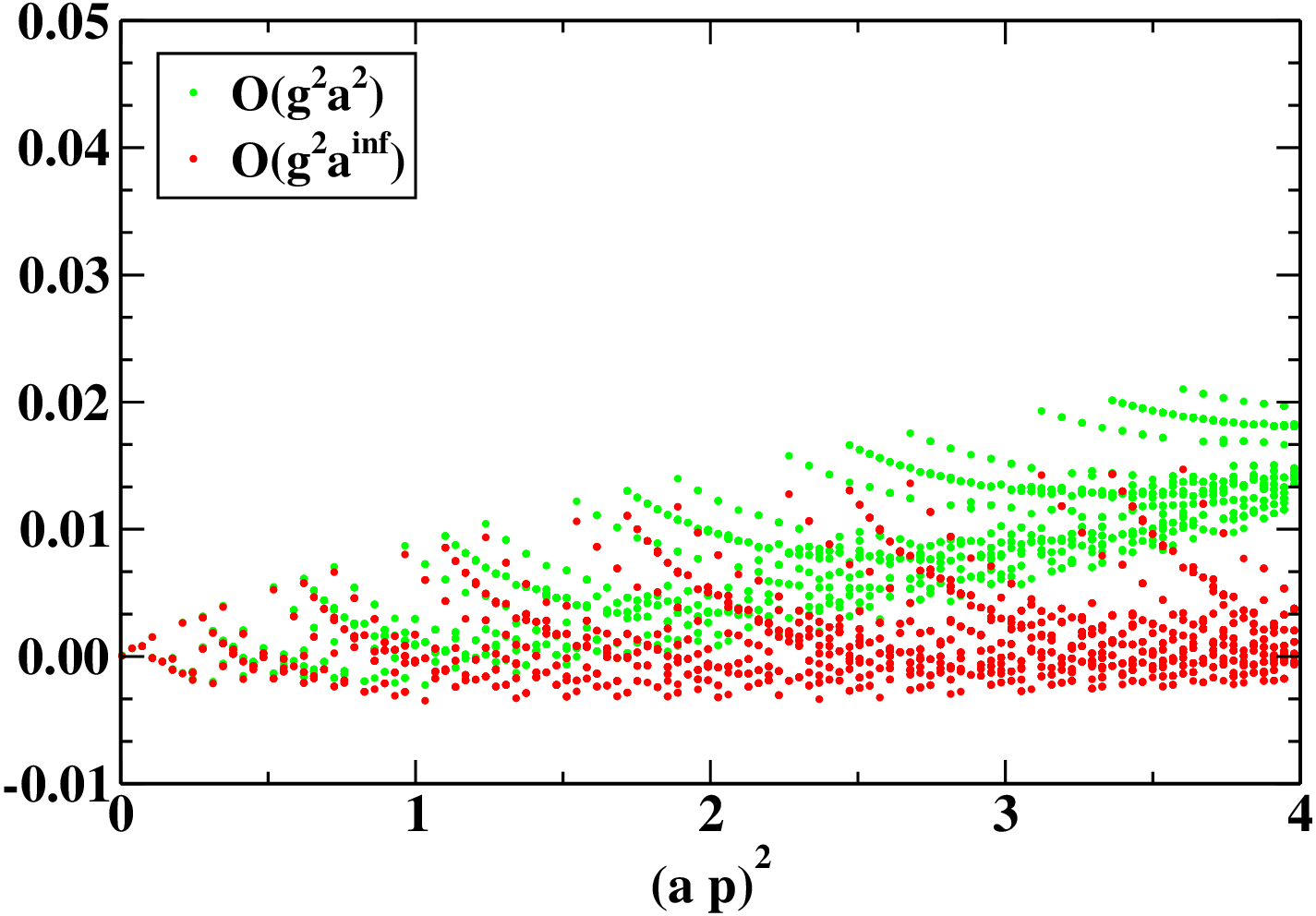}}
\caption{$D{\cal Z}_P(a,p)/g^2$ as a function of $(a\,p)^2$. The notation is the same as for Fig.~\ref{Zq_pert}.}
\label{ZP_pert}
\end{figure}

\begin{figure}[!h]
{\includegraphics[scale=0.345]{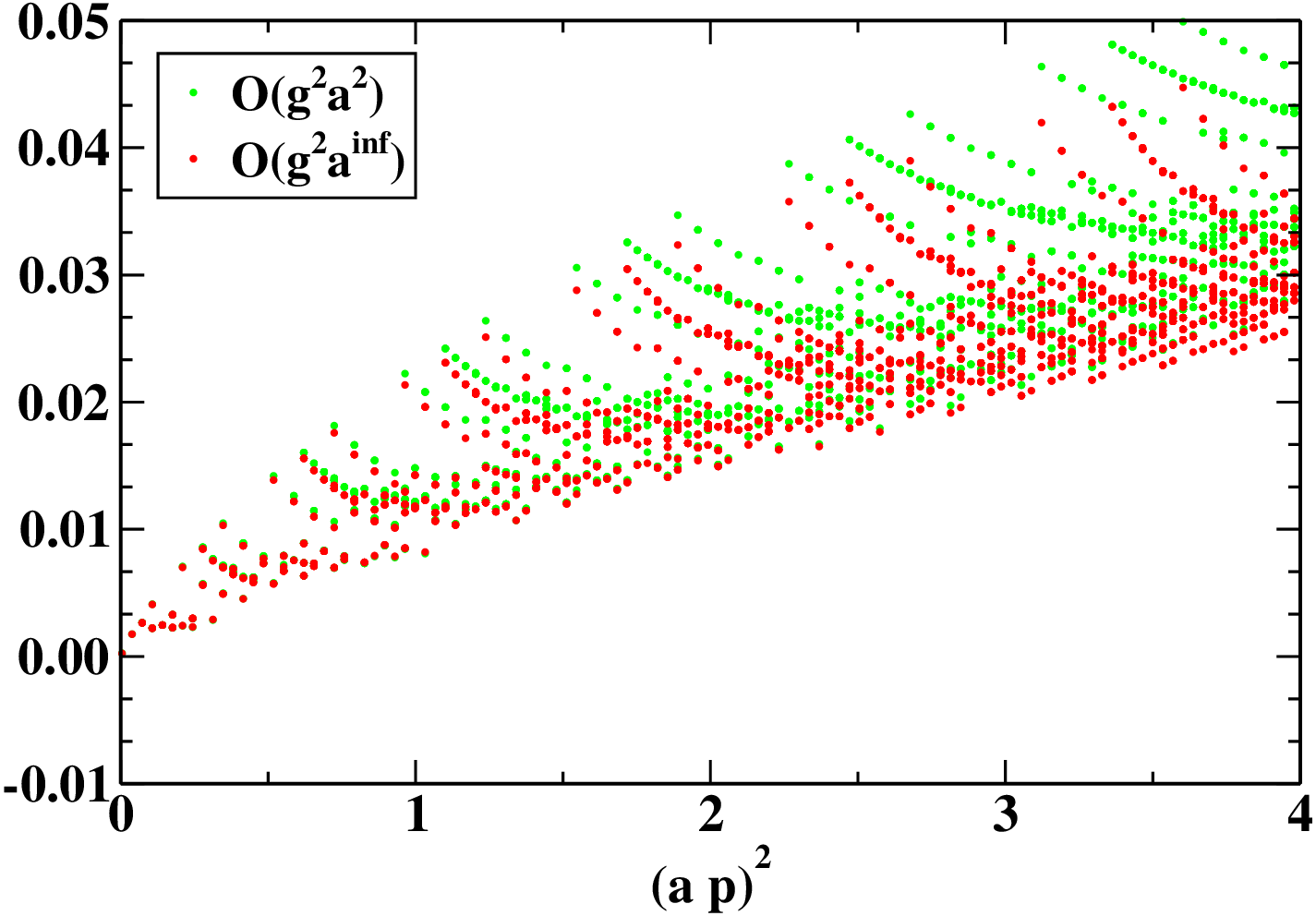}}$\quad$
{\includegraphics[scale=0.345]{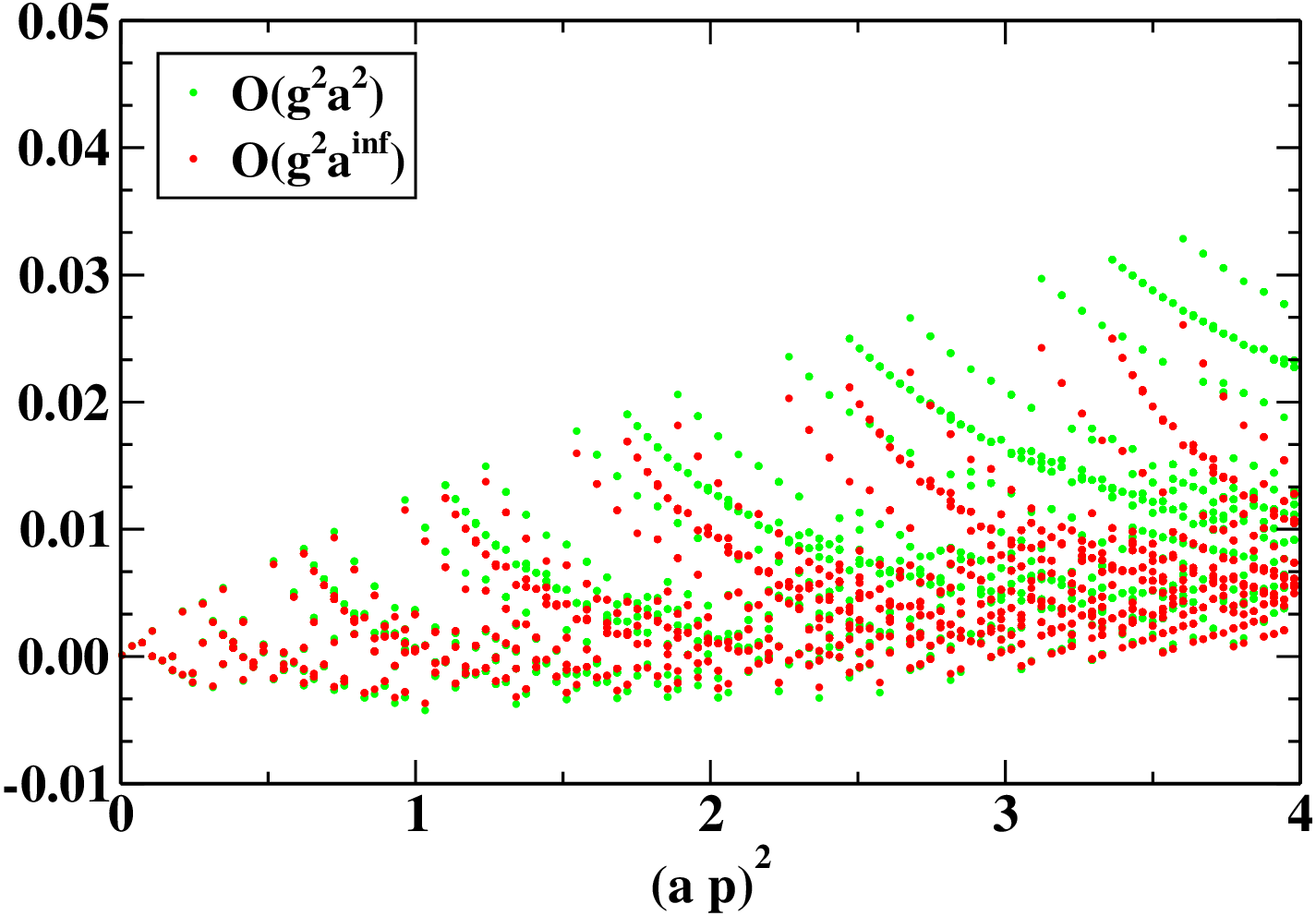}}
\caption{$D{\cal Z}_V(a,p)/g^2$ as a function of $(a\,p)^2$. The notation is the same as for Fig.~\ref{Zq_pert}.}
\label{ZV_pert}
\end{figure}

In Figs.~\ref{Zq_pert} - \ref{ZT_pert} we plot $D{\cal Z}_q(a,p)/g^2$ and $D{\cal
Z}_{\cal O}(a,p)/g^2$ for the ultra-local bilinears corresponding to
the Iwasaki improved action using a lattice size of $24^3\times48$ and
several values of $(a\,p)^2$ within the range of 0-4. For comparison
we also include the corresponding ${\cal O}(g^2\,a^0)/g^2$ terms
computed in Ref.~\cite{Constantinou:2009tr}. Since a clover term is
included in the $N_f{=}2$ ensembles, we consider both values: $\csw=0$
(left figures) and $\csw=1$ (right figures). An immediate observation is
that momenta with the same $(a\,p)^2$ lead to different lattice
artifacts, which is expected, since beyond ${\cal O}(a^0)$ these terms
depend not only on the length, but also on the direction of the
four-vector $p$, due to the presence of Lorentz noninvariant
structures, such as:
\be
a^2\,\frac{p4}{p2} \equiv \frac{\sum_\rho a^4 p^4_\rho}{\sum_\rho a^2 p^2_\rho}\,,
\label{p4p2}
\ee
appearing to ${\cal O}(g^2 a^2)$.
It is also interesting to see that the lattice artifacts depend on the
operator under study in a non-predictible way since in some cases
($\Zq$, $\ZV$, $\ZT$) the inclusion of a clover term diminishes the
artifacts, while in another ($\ZS$) it enhances them. For the case of
$\ZA$ and $\ZP$ the lattice artifacts for $\csw=0,1$ are comparable.
As expected, comparison between ${\cal O}(g^2\,a^2)$ and ${\cal O}(g^2\,a^\infty)$
for small values of the momenta ($(a\,p)^2\ll 1$) reveals a very good
agreement, since the ${\cal O}(g^2\,a^2)$ terms are the
leading contributions of the lattice artifacts. For larger momenta the
difference between ${\cal O}(g^2\,a^2)$ and ${\cal O}(g^2\,a^\infty)$
is more apparent, as will be discussed in Section~\ref{sec5}.

\FloatBarrier
\vskip 0.3cm

\FloatBarrier
\vskip 0.3cm

\FloatBarrier
\vskip 0.3cm

\FloatBarrier
\vskip 0.3cm

\begin{figure}[!h]
{\includegraphics[scale=0.345]{./ZA_csw_0.eps}}$\quad$
{\includegraphics[scale=0.345]{./ZA_csw_1.eps}}
\caption{$D{\cal Z}_A(a,p)/g^2$ as a function of $(a\,p)^2$. The notation is the same as for Fig.~\ref{Zq_pert}.}
\label{ZA_pert}
\end{figure}
\FloatBarrier
\vskip 0.3cm

\begin{figure}[!h]
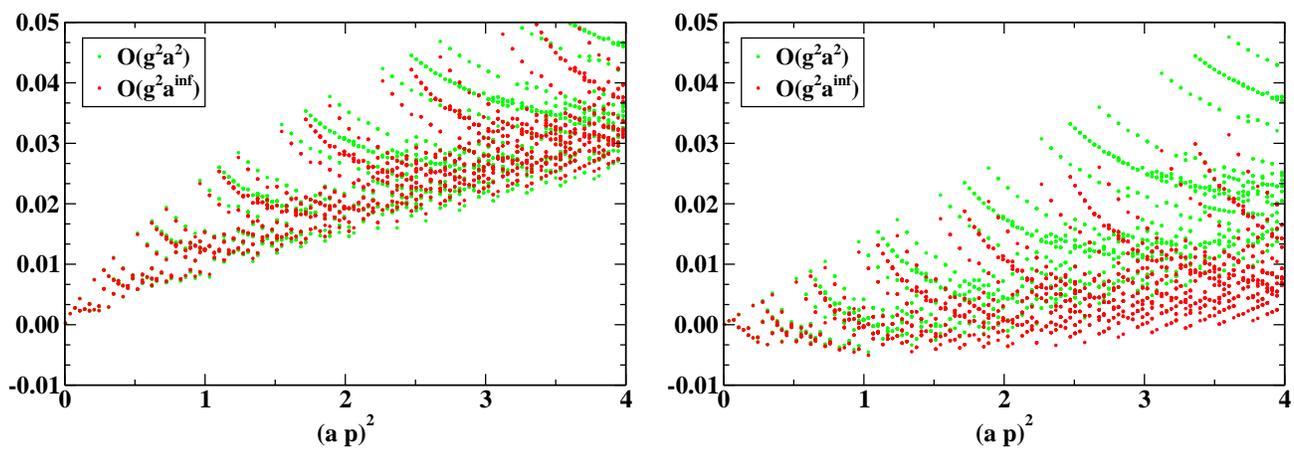

{\includegraphics[scale=0.345]{./Zt_csw_0.eps}}$\quad$
{\includegraphics[scale=0.345]{./Zt_csw_1.eps}}
\caption{$D{\cal Z}_T(a,p)/g^2$ as a function of $(a\,p)^2$. The notation is the same as for Fig.~\ref{Zq_pert}.}
\label{ZT_pert}
\end{figure}

\section{Non-perturbative evaluation}
\label{sec5}

\subsection{Chiral extrapolation}
\label{sec5.1}

\begin{figure}[!h]
\centerline
{\includegraphics[scale=0.335]{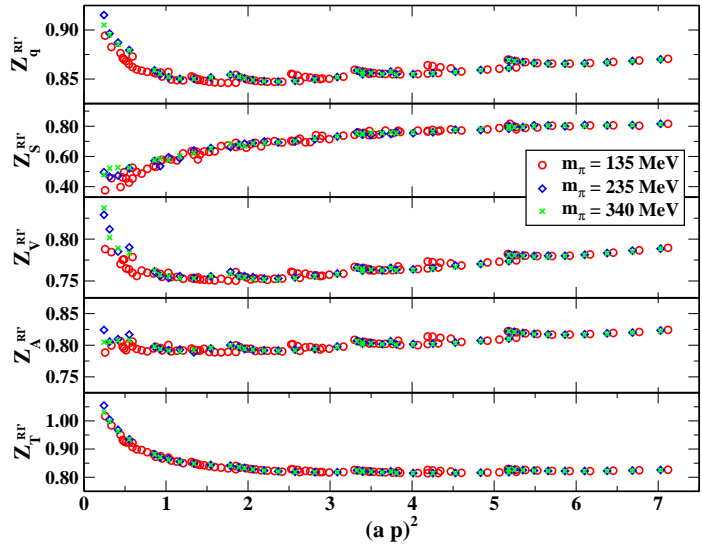}}
\caption{Pion mass dependence of $\Zq$ and the RFs of the ultra-local
bilinears for  $N_f{=}2$ at $\beta{=}2.10$ as a function of the
renormalization scale.}
\label{Zlocmpi}
\end{figure}

\begin{figure}[!h]
\centerline
{\includegraphics[scale=0.335]{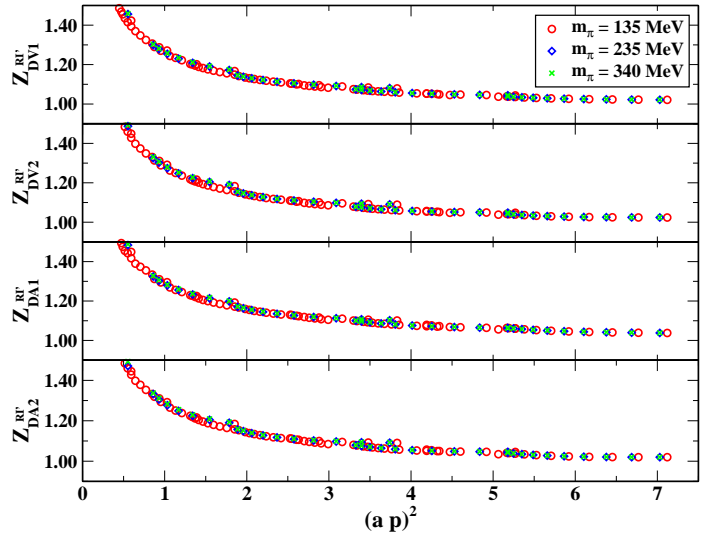}}
\caption{Pion mass dependence of the RFs of the one-derivative
vector and axial operators for $N_f{=}2$ at $\beta{=}2.10$ as a
function of the renormalization scale.}
\label{Z1Dmpi}
\end{figure}

\begin{figure}[!h]
\centerline
{\includegraphics[scale=0.335]{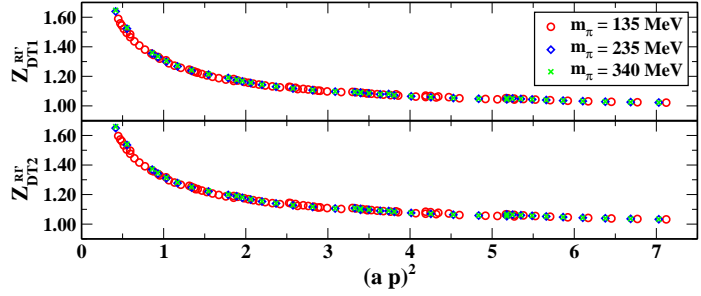}}
\caption{Pion mass dependence of the RFs of the one-derivative
tensor operator for $N_f{=}2$ at $\beta{=}2.10$ as a function of the
renormalization scale.}
\label{Z1DTmpi}
\end{figure}

\begin{figure}[!h]
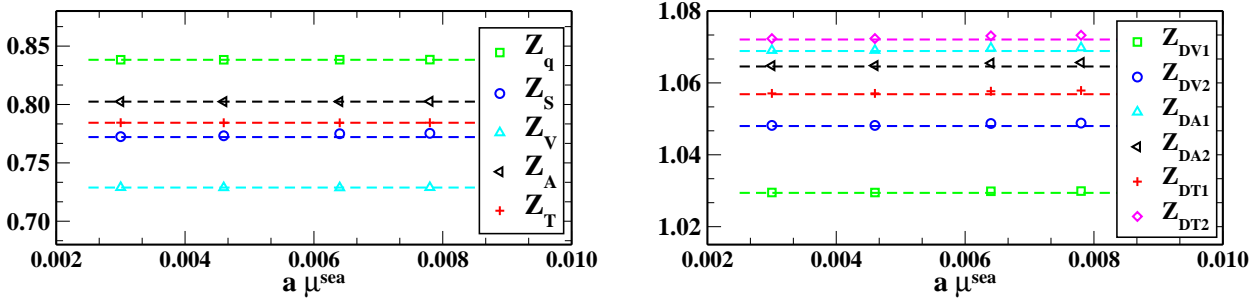

\centerline{\includegraphics[scale=0.42]{./Z_local_Nf4_b2.1_vs_mpi_a2p2_3.009_RI.eps}$\qquad$
{\includegraphics[scale=0.42]{./Z_oneD_Nf4_b2.1_vs_mpi_a2p2_3.009_RI.eps}}}
\caption{Pion mass dependence of $\Zq$ and the RFs of the ultra-local
bilinears (left plot) and the one-derivative operators (right plot)
for $N_f{=}4$ at $\beta{=}2.10$ as a function of $a\,\mu^{\rm sea}$ at
$(a\,p)^2=3$.}
\label{Zloc1Dmpi2}
\end{figure}

In order to obtain the renormalization functions in the chiral limit
we perform an extrapolation using a linear fit with respect to
$m_\pi^2$. We find that the RFs obtained in this work have a very mild 
dependence on the pion mass for all ensembles. In fact, with the exception
of a few small values of $(a\,p)^2$, there is no visible pion mass dependence
within the small statistical errors. Allowing a slope and performing a
linear extrapolation with respect to $m_\pi^2$ the data yield a slope
consistent with zero. Figs.~\ref{Zlocmpi} - \ref{Zloc1Dmpi2} demonstrate
the pion mass dependence of the RFs using the $N_f{=}2$ and $N_f{=}4$
ensembles at $\beta{=}2.10$. The statistical errors are too small to
be visible. Figs.~\ref{Zlocmpi} - \ref{Z1DTmpi} provide a more general
picture of the $m_\pi$-dependence by displaying the RFs as a function
of the renormalization scale ($\mu^2=p^2$), while the two plots of
Fig.~\ref{Zloc1Dmpi2} show the data at $(a\,p)^2=3$ as a function of
the twisted mass $a\,\mu^{\rm sea}$. These plots show clearly that the
slope of the fit is consistent with zero.

In this discussion the renormalization function of the pseudoscalar
density, $\ZP$, has been excluded since there is pion pole
contamination that needs to be taken into account. Thus, a polynomial
fit with respect to the pion mass is not suitable. An approriate
chiral extrapolation of $\ZP$ and the ratio $\ZP/\ZS$ is discussed in
the following Subsection.

\FloatBarrier

\subsection{Pion-pole subtraction of $\ZP$ and $\ZP/\ZS$}

The correlation functions of the pseudoscalar operator have pion-pole
contamination which needs to be treated carefully. In order to
subtract the pole contribution we use a 2- and 3-parameter Ansatz for
the pseudoscalar amputated vertex function, $\Gamma_P$, of the form:
\bea
\label{PionPole}
F^{(2)}_P &=& a_P + \frac{c_P}{m_\pi^2}\,,\\
F^{(3)}_P &=& a_P + b_P\,m_\pi^2 + \frac{c_P}{m_\pi^2}\,.
\eea
The fit parameters depend on both the momentum and the
value of $\beta$ (i.e. $a_P \equiv a_P(\beta,p)$), and thus we estimate
them separately on each value of $p$ and $\beta$. We find that the
coefficient $b_P$ is very small and competes with $c_P$ in the
3-parameter fit. In addition, they both carry large statistical
errors, which result in a large error in the final determination of $\ZP$ once the
term $c_{P}/m_\pi^2$ is subtracted from the pseudoscalar matrix elements:
\be
\Gamma_P^{\rm sub} = \Gamma_P - \frac{c_{P}}{m_\pi^2}.
\label{polesub}
\ee 

\begin{figure}[!h]
{\includegraphics[scale=0.61]{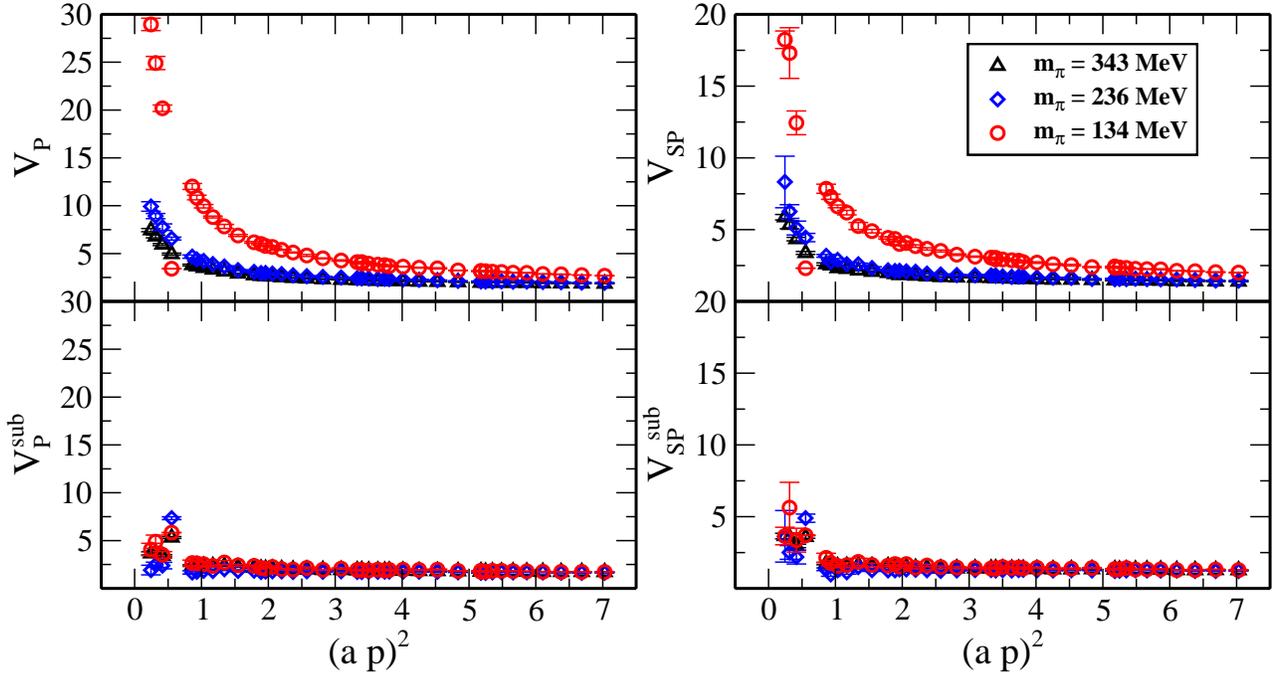}}
\caption{Upper panels: Eq.~(\ref{fit_ratio}) (left plot) and Eq.~(\ref{fit_ratio2})
(right plot) for $N_f{=}2$ at $\beta{=}2.10$ as a function of $(a\,p)^2$
  for each value of the pion mass before the pole subtraction. Lower
  panels: Eq.~(\ref{fit_ratio}) (left plot) and Eq.~(\ref{fit_ratio2})
(right plot) for $N_f{=}2$ at $\beta{=}2.10$ as a function of $(a\,p)^2$
  for each value of the pion mass after the pole subtraction.}
\label{fig3}
\end{figure}

\begin{figure}[!h]
{\includegraphics[scale=0.61]{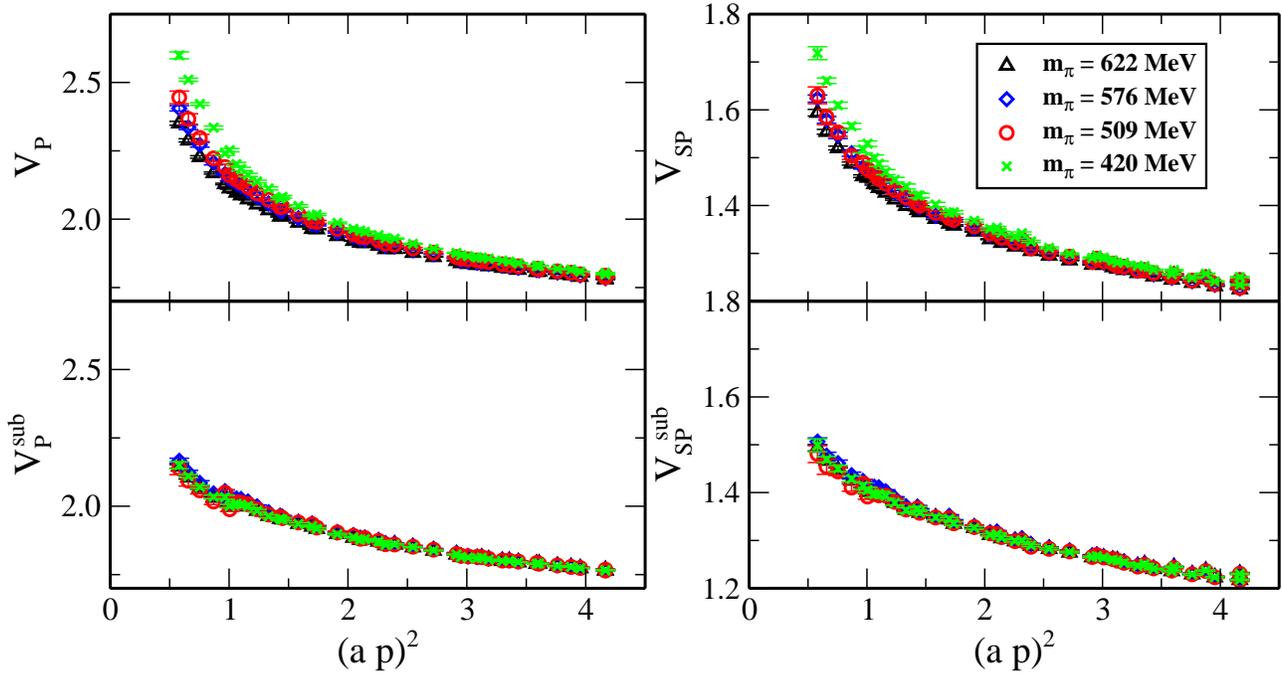}}
\caption{As Fig.~\ref{fig3} but for the $N_f{=}4$ and $\beta{=}2.10$ ensembles.}
\label{fig4}
\end{figure}

A way around this problem is to employ the 2-parameter fit of
Eq.~(\ref{PionPole}) directly to the ratio:
\be
V_P(p,\,m_\pi) =
\frac{\Gamma_P(p,\,m_\pi)}{\Zq(p,\,m_\pi)\,
C_P^{\mbox{\scriptsize RI$^{\prime}$},\,{\overline{\rm MS}}} (p,2\, \rm{GeV})}\,,
\label{fit_ratio}
\ee
where $C_P^{\mbox{\scriptsize RI$^{\prime}$},\,{\overline{\rm MS}}}(p,2\, \rm{GeV})$
is the conversion to the ${\overline{\rm MS}}$ scheme and the
evolution to a scale of 2 GeV. This fit allows us to obtain 
directly $\ZP^{\overline{\rm MS}}$ in the ${\overline{\rm MS}}$-scheme
and at the chiral limit~\footnote{Alternative fit
functions and their stability are discussed in
Ref.~\cite{Constantinou:2014fka}.} from $1/a_{P}$.
In a similar manner, we obtain directly $\ZS/\ZP$ from the
coefficient $a_P$ computed from the 2-parameter fit of
\begin{equation}
V_{SP}(p,m_\pi) = \frac{\Gamma_P(p,m_\pi)}{\Gamma_S(p,m_\pi)} \,.
\label{fit_ratio2}
\end{equation}
As an example of the pion pole contamination and its subtraction we
show, in Fig.~\ref{fig3}, $V_P(p,\,m_\pi)$  and
$V_{SP}(p,\,m_\pi)$  using the $N_f{=}2$ ensembles at
$\beta{=}2.10$ for each value of the pion mass before
and after the sutraction of the pion pole term,
$\frac{c_P(p^2)}{m_\pi^2}$. Fig.~\ref{fig4} is similar
to Fig.~\ref{fig3} for the $N_f{=}4$ and $\beta{=}2.10$ ensembles. The
range of the y-axis is the same for the unsubtracted and subtracted
cases in order to see clearly the effectiveness of the pion-pole
subtraction. 
\FloatBarrier
\FloatBarrier

\subsection{${\overline{\rm MS}}$-scheme}

In order to compare lattice values to experimental results one must
convert to a universal renormalization scheme and use a reference
renormalization scale. Typically one chooses the ${\overline{\rm MS}}$-scheme 
at a scale $\mu$ of 2 GeV. For the conversion from the RI$'$ to the
${\overline{\rm MS}}$ scheme we use the intermediate Renormalization
Group Invariant (RGI) scheme, which is scale independent and
relates the RI$'$ and ${\overline{\rm MS}}$ results are follows:
\be
Z^{\rm RGI}_{\cal O} =
Z_{\cal O}^{\mbox{\scriptsize RI$^{\prime}$}} (\mu) \, 
\Delta Z_{\cal O}^{\mbox{\scriptsize RI$^{\prime}$}}(\mu) =
Z_{\cal O}^{\overline{\rm MS}} (2\,{\rm GeV}) \,
\Delta Z_{\cal O}^{\overline{\rm MS}} (2\,{\rm GeV})\,.
\ee
The conversion factor can be read from the above relation:
\be
Z_{\cal O}^{\overline{\rm MS}} (2\,{\rm GeV})  \equiv 
C_{\cal O}^{{\rm RI}',{\overline{\rm MS}}}(\mu,2\,{\rm GeV})\,
Z_{\cal O}^{{\rm RI}'} (\mu)\,,\qquad
C_{\cal O}^{{\rm RI}',{\overline{\rm MS}}}(\mu,2\,{\rm GeV})=
\frac{\Delta Z_{\cal O}^{\mbox{\scriptsize RI$^{\prime}$}}(\mu)}
     {\Delta Z_{\cal O}^{\overline{\rm MS}}(2\,{\rm  GeV})}\,,
\ee
where the scheme dependent quantity $\Delta Z_{\cal O}^{\mathcal S}(\mu)$ 
can be expressed in terms of the $\beta$-function and the anomalous
dimension, $\gamma_{\cal O}^S \equiv \gamma^S$ of the operator 
${\cal O}$ (for definitions see Appendix~\ref{appA}): 
\be
\Delta Z_{\cal O}^{\mathcal S} (\mu) =
  \left( 2 \beta_0 \frac {{g^{\mathcal S} (\mu)}^2}{16 \pi^2}\right)
^{-\frac{\gamma_0}{2 \beta_0}}
 \exp \left \{ \int_0^{g^{\mathcal S} (\mu)} \! \mathrm d g'
  \left( \frac{\gamma^{\mathcal S}(g')}{\beta^{\mathcal S} (g')}
   + \frac{\gamma_0}{\beta_0 \, g'} \right) \right \}\,.
\ee
We employ the 3-loop approximation which simplifies to:
\bea
\label{DZ_RGI}
\Delta Z_{\cal O}^{\mathcal S} (\mu) &\hspace{-0.1cm}=\hspace{-0.1cm}&
  \left( 2 \beta_0 \frac {{g^{\mathcal S} (\mu)}^2}{16 \pi^2}\right)
^{-\frac{\gamma_0}{2 \beta_0}}
\Bigg( 1  + \frac {g^{\mathcal S} (\mu)^2}{16 \pi^2}\,
\frac{\beta_1 \gamma_0-\beta_0 \gamma^S_1}{2 \beta_0^2} +  \\
&& 
\frac {{g^{\mathcal S} (\mu)}^4}{(16 \pi^2)^2}\,
   \frac{-2 \beta_0^3 \gamma^S_2+\beta_0^2 (\gamma^S_1 (2
    \beta_1+\gamma^S_1)+2 \beta_2 \gamma_0)-2 \beta_0 \beta_1
    \gamma_0 (\beta_1+\gamma^S_1)+\beta_1^2 \gamma_0^2}{8
     \beta_0^4}  \Bigg)\,, \nonumber
\eea
where the coupling constant, $g^S(\mu)$, is needed in both the
${\overline{\rm MS}}$ and RI$'$ schemes; their expressions coincide to
three loops and read~\cite{Alekseev:2002zn} 
\footnote{Sign differences in some terms of Eq.~(\ref{gS}) compared to
Ref.~\cite{Alekseev:2002zn} are related to alternative definition of
the $\beta$-function}:
\bea
\label{gS}
\frac{{g^{{\overline{\rm MS}},{\mbox{\scriptsize
          RI$^{\prime}$}}}(\mu)}^2}{16\pi^2}\Big{|}_{\rm 3-loop} = 
\frac{1}{\beta_0\, L} 
 - \frac{\beta_1}{\beta_0^3} 
\frac{\log L}
{L^2} 
+\frac{1}{\beta_0^5}
\frac{\beta_1^2 \log^2 L
- \beta_1^2 \log L  +
\beta_2 \beta_0 - \beta_1^2}{L^3})
\,,\quad L = \log \frac{\mu^2}{\Lambda^2_{\overline{\rm MS}}}\,.
\eea

For $\Lambda_{\overline{\rm MS}}$ we employ the value 315 MeV and 296 MeV
for $N_f{=}2$~\cite{Jansen:2011vv,Fritzsch:2012wq} and $N_f{=}4$~\cite{PDG14}, 
respectively.

\section{Results}

In this section, we present our results for the renormalization
functions in the ${\overline{\rm MS}}$ scheme at a scale of 2~GeV. The
final data correspond to the non-perturbative values after subtracting
the lattice artifacts to ${\cal O}(g^2\,a^\infty)$. Although, with the
exception of $\ZP$, the dependence of the RFs on the pion mass is not
significant, we nevertheless perform a chiral extrapolation of the RFs
using data at the same $\beta$ and $N_f$ ensembles obtained for
different pion masses as discussed in subsection~\ref{sec5.1}. 

As can be seen in Figs.~\ref{ZavALL} - \ref{ZtD} the 
${\cal O}(g^2\,a^\infty)$-subtracted RFs (magenta diamond points) have
a mild dependence on $(ap)^2$, which is removed by extrapolating to
zero, using the Ansatz 
\begin{equation}
Z_{\cal O}(a\,p) = Z_{\cal O}^{(0)} + Z_{\cal O}^{(1)}\cdot(a\,p)^2\,,
\label{Zfinal}
\end{equation}
where $ Z_{\cal O}^{(0)}$ corresponds to our final value on the
renormalization functions. To extract the RFs reliably one needs to
consider momenta in the range $\Lambda_{QCD}<p<1/a$. We relax the
upper bound to be $\sim 4/a$ to $7/a$, which is justified by the
weak dependence  of our results on $(a\,p)^2$. Therefore, for each
value of $\beta$ we consider momenta $(a\,p)^2 \ge 2$ for which
perturbation theory is trustworthy and lattice artifacts are still
small enough. 

\begin{figure}[!h]
{\includegraphics[scale=0.55]{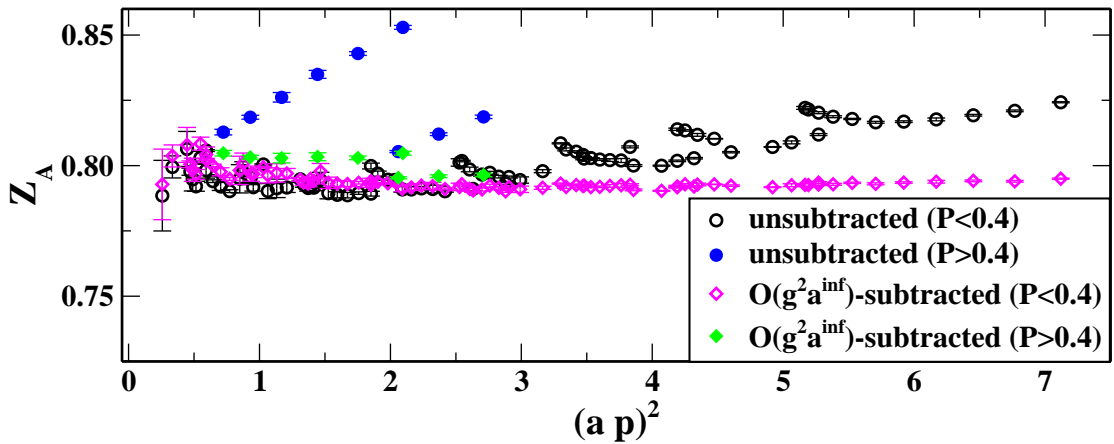}}
\caption{$\ZA$ as a function of the renormalization scale. Filled blue
circles and filled green diamonds correspond to unsubtracted and
${\cal O}(g^2\,a^\infty)$-subtracted data using momenta with $P\equiv p4/p2>0.4$.} 
\label{Zab2.1csw}
\end{figure}

\begin{figure}[!h]
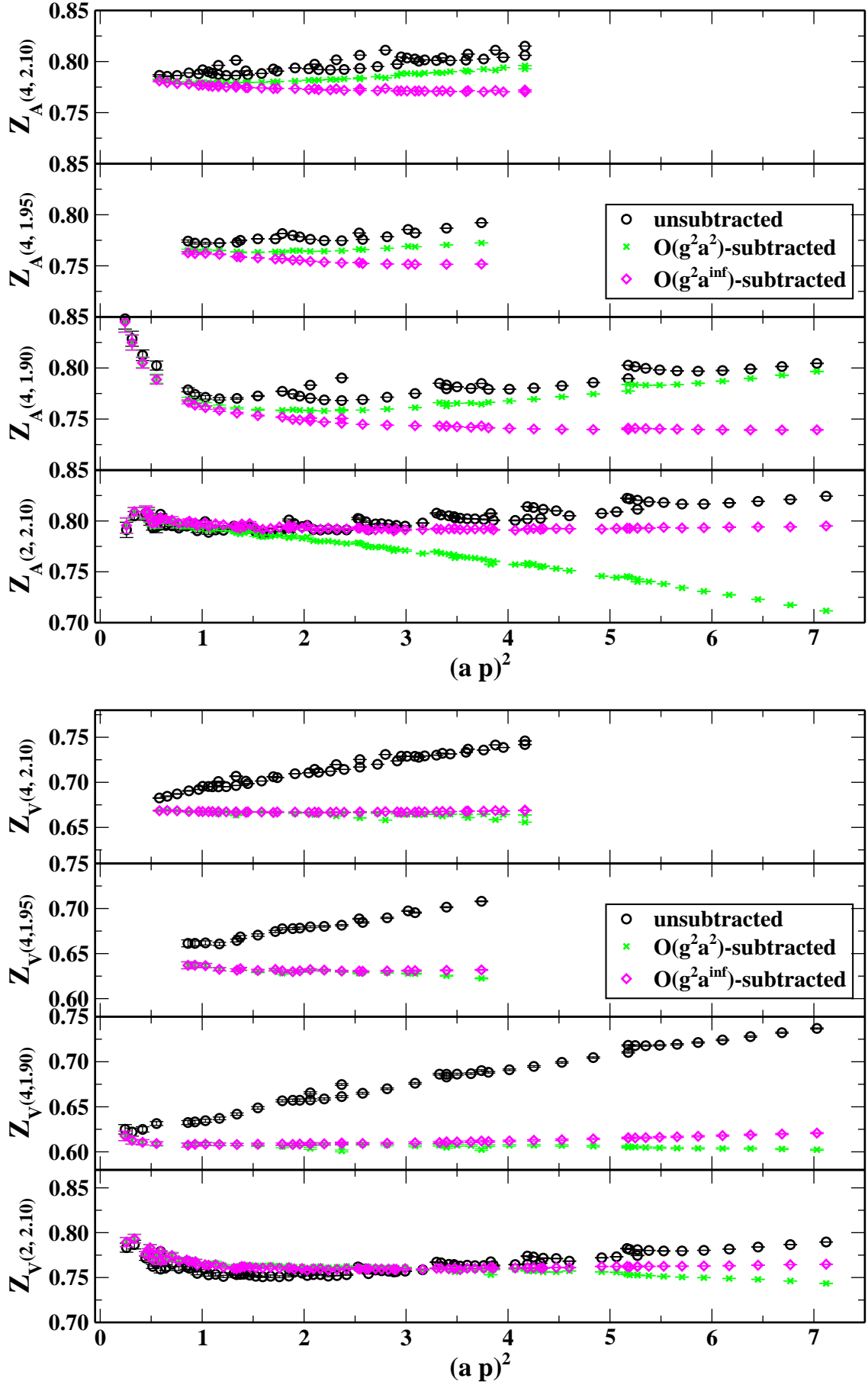

{\includegraphics[scale=0.55]{./Za_Nf2_csw_Nf4.eps}}\\[3ex]
{\includegraphics[scale=0.55]{./Zv_Nf2_csw_Nf4.eps}}
\caption{$\ZA$ (upper) and $\ZV$ (lower) as a function of the
renormalization scale. The arguments $a,\,b$ of $Z(a,b)$ correspond to
$N_f$ and $\beta$, respectively. From top to bottom, the data
correspond to increasing the lattice spacing. Non-perturbative values
are shown with the black circles, ${\cal O}(g^2\,a^2)$-subtracted
with the green crosses and ${\cal O}(g^2\,a^\infty)$-subtracted 
with magenta diamonds.}
\label{ZavALL}
\end{figure}

\begin{figure}[!h]
{\includegraphics[scale=0.55]{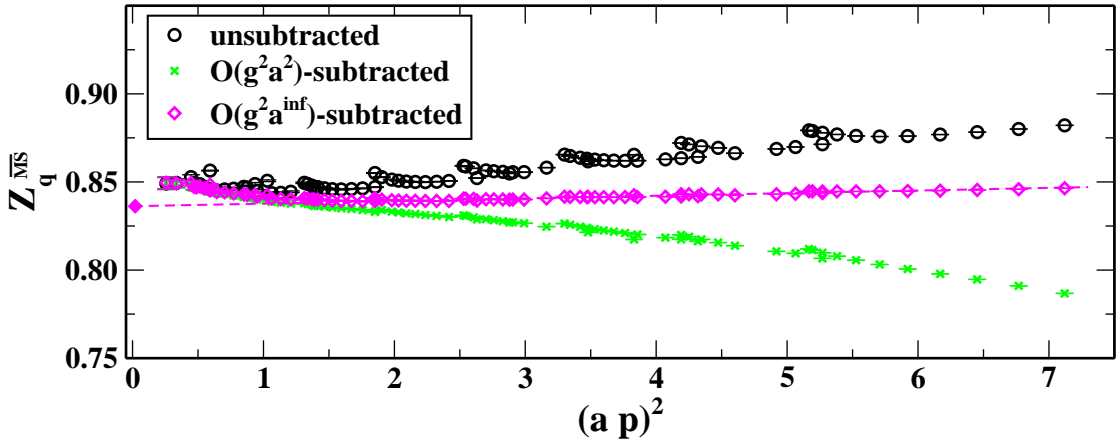}}
\vspace{-0.15cm}
\caption{Renormalization of the fermion field for $N_f{=}2$ twisted
  mass clover-improved fermions. The data correspond to the
${\overline{\rm MS}}$ scheme at a reference scale of 2 GeV and are
plotted against the initial renormalization scale, $(a\,\mu)^2=(a\,p)^2$. 
Black circles (magenta diamonds, green crossed) denote the unsubtracted
(${\cal O}(g^2\,a^\infty)$-subtracted, ${\cal O}(g^2\,a^2)$-subtracted)
non-perturbative data.}
\label{Zq} 
\end{figure}

\begin{figure}[!h]
{\includegraphics[scale=0.55]{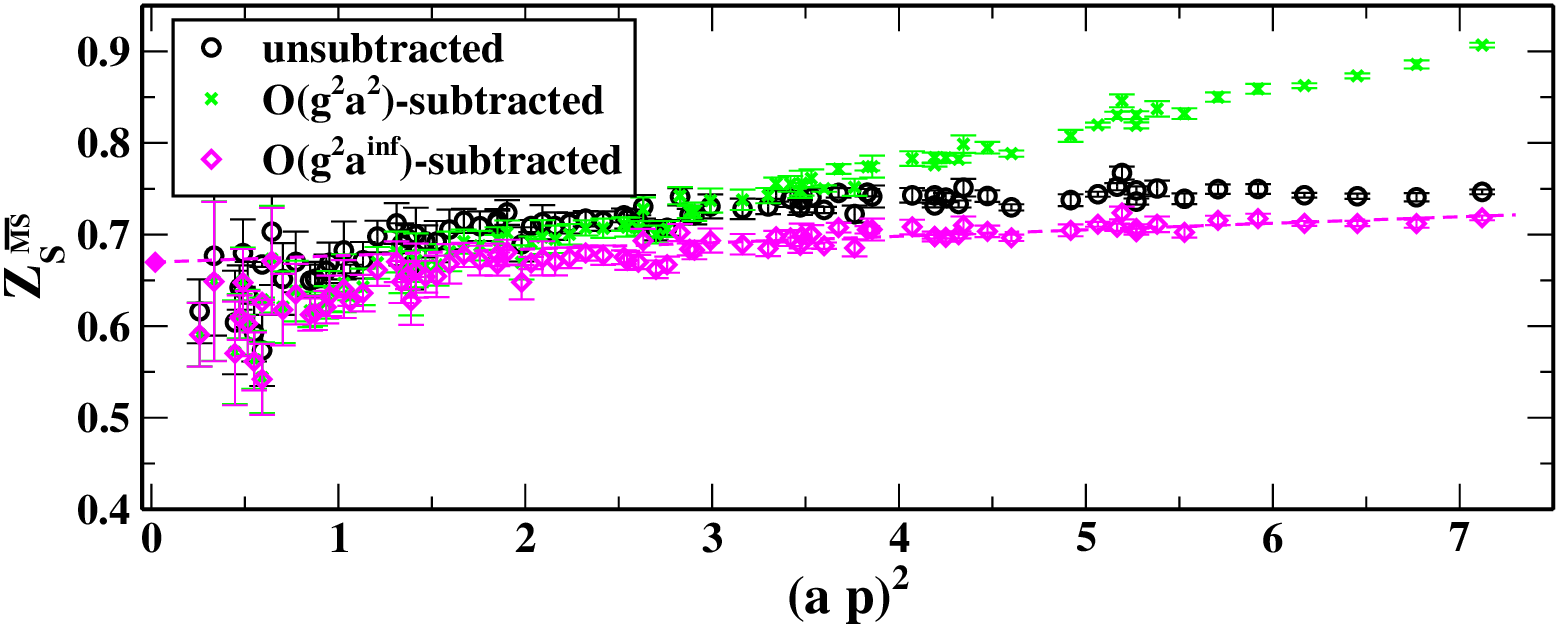}}
\vspace{-0.15cm}
\caption{ The renormalization function of the scalar operator. The
notation is the same as that of Fig.~\ref{Zq}.}
\label{Zs} 
\end{figure}

\begin{figure}[!h]
{\includegraphics[scale=0.55]{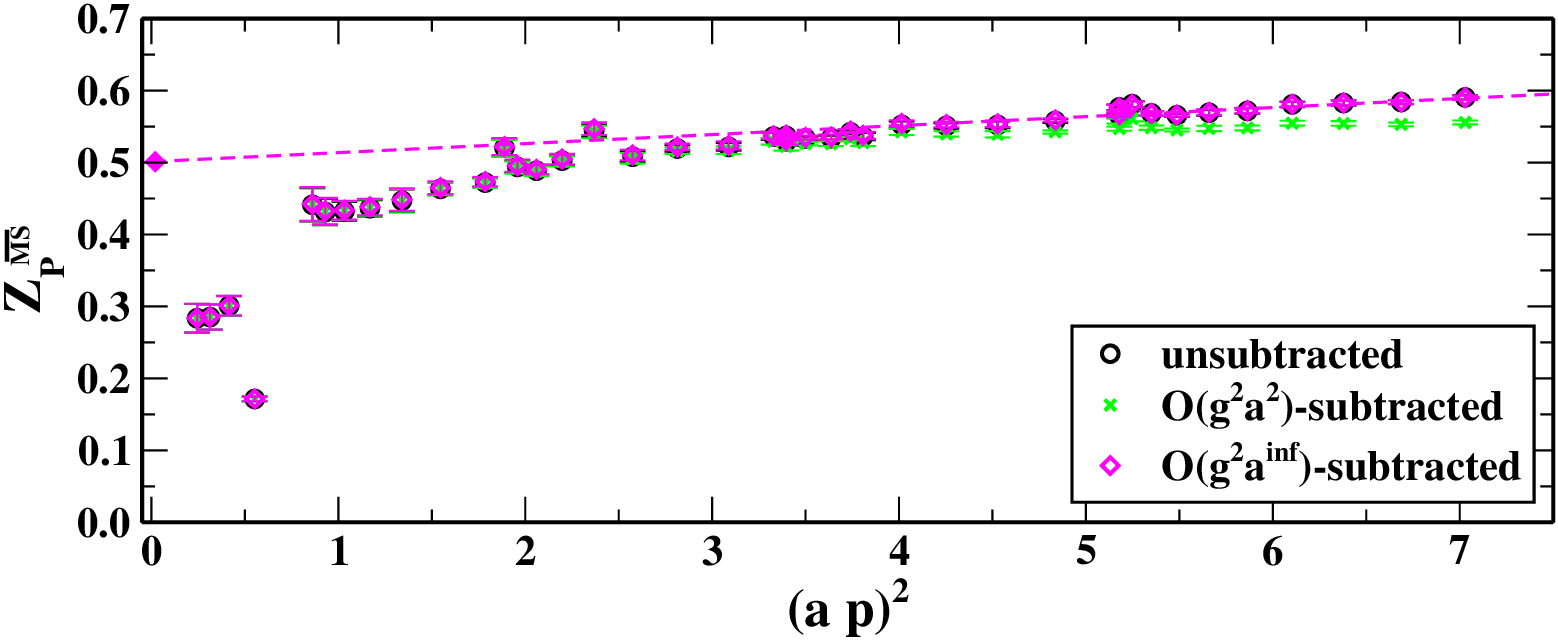}}
\caption{ $\ZP$ after removal of the pion-pole term. The notation is the
same as that of Fig.~\ref{Zq}.}
\label{Zp} 
\end{figure}

\begin{figure}[!h]
{\includegraphics[scale=0.55]{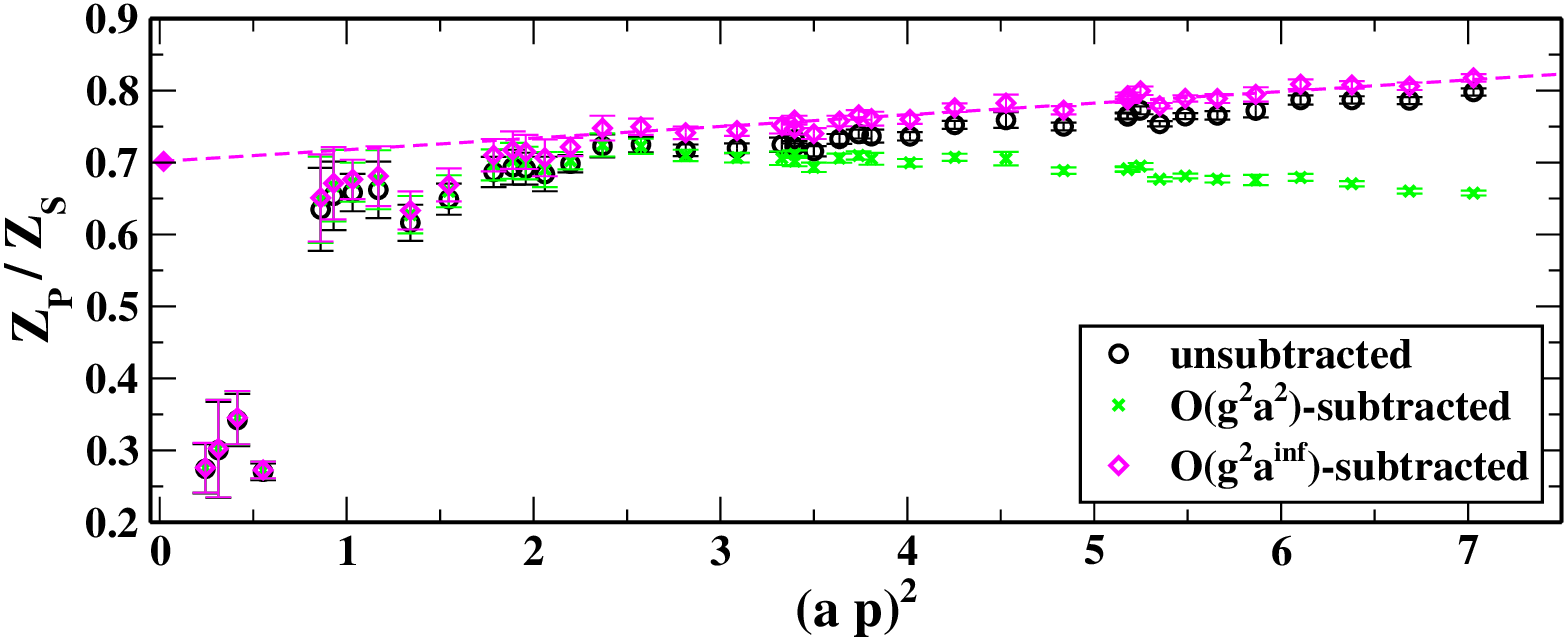}}
\caption{The ratio $\ZP/\ZS$ after removal of the pion-pole term. The
  notation is the same as that of Fig.~\ref{Zq}.}
\label{ZpoZs} 
\end{figure}

\begin{figure}[!h]
{\includegraphics[scale=0.55]{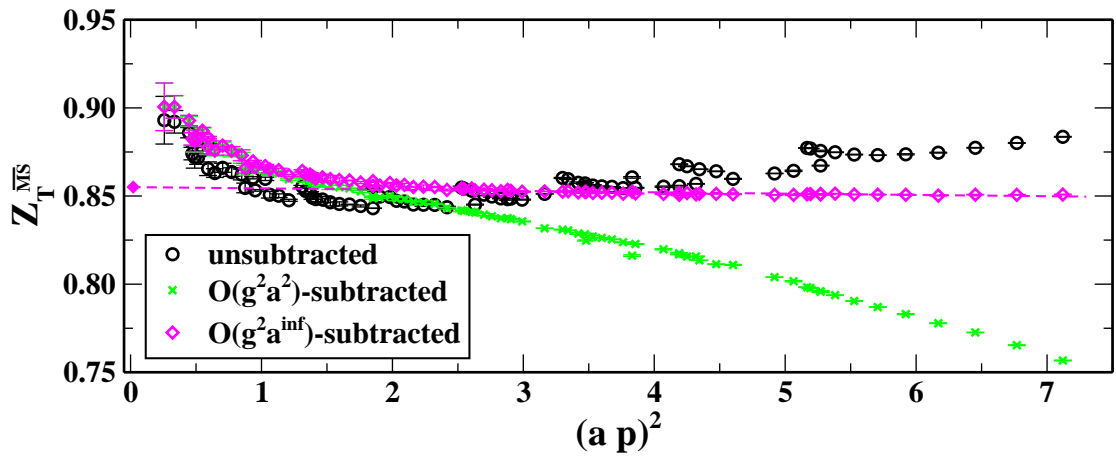}}
\caption{The renormalization function of the tensor operator. The
notation is the same as that of Fig.~\ref{Zq}.}
\label{Zt} 
\end{figure}

\begin{figure}[!h]
{\includegraphics[scale=0.55]{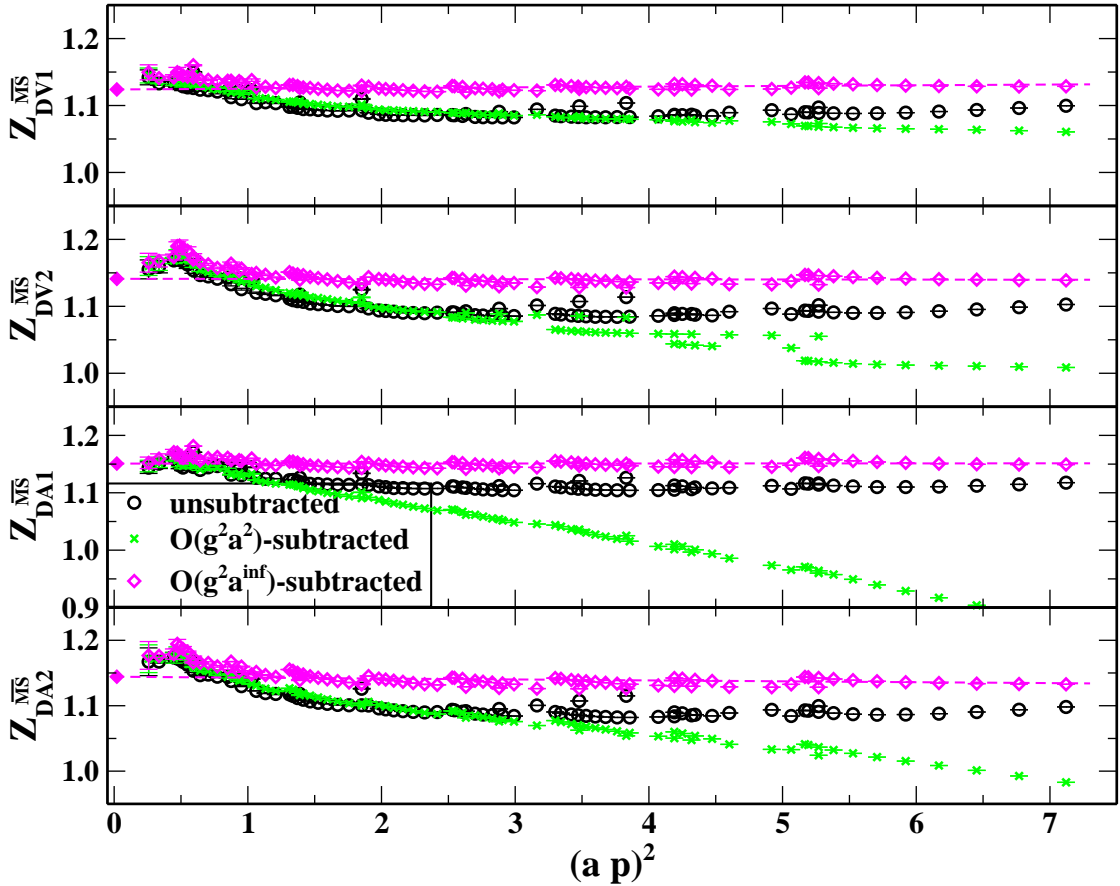}}
\vspace{-0.15cm}
\caption{The renormalization functions of the one-derivative vector
  and axial operators. The notation is the same as that of Fig.~\ref{Zq}.}
\label{ZvDaD} 
\end{figure}

\begin{figure}[!h]
{\includegraphics[scale=0.55]{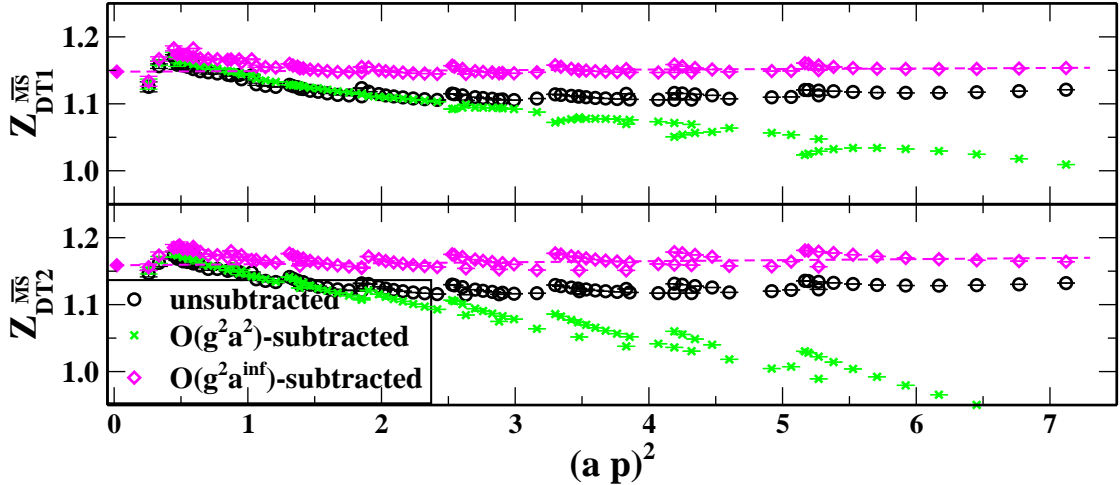}}
\vspace{-0.15cm}
\caption{ The renormalization functions of the one-derivative tensor
  operators. The notation is the same as that of Fig.~\ref{Zq}.}
\label{ZtD} 
\end{figure}

From our analysis we find that the data for the RFs depend not only
on $(a\,p)^2$, but also on the directions of the momentum. Noting
that, for democratic momenta (in all directions, not only spatial) the
value of $p4/p2^2$ equals $0.25$, we find empirically that data
produced on momenta with the ratio of Eq.~(\ref{p4p2}), $p4/p2^2$,
being $ > 0.4$, have a behavior that deviates from the general
$(a\,p)^2$ curve. The choice of such a momentum ratio as a 
criterion is justified by the fact that such Lorentz non-invariant
contributions appear in the perturbative computation at higher orders
in the lattice spacing (e.g., $p4/p2$ for ${\cal O}(a^2)$). 
Thus, high values of this ratio are an indication of large lattice
artifacts from higher loops. As an example, we demonstrate $\ZA$ in
Fig.~\ref{Zab2.1csw} including the data obtained at momenta satisfying
$P\equiv p4/p2 > 0.4$. These are shown by the filled blue circles and
filled green diamonds corresponding to unsubtracted and 
${\cal O}(g^2\,a^\infty)$-subtracted data, respectively. As can be seen, 
 the filled symbols have different behavior than the open
symbols. Although the subtraction of one-loop lattice artifacts
reduces the difference, the higher order artifacts are not
negligible. The data for these momenta  have been excluded from
the final analysis of all RFs. A similar study is presented in
Refs.~\cite{Alexandrou:2010me,Alexandrou:2012mt}.

\FloatBarrier

While statistical errors are very small, a careful investigation of
systematic errors is required. A small systematic effect comes from
the asymmetry of our lattices, both because they are larger in their
time extent and because of the antiperiodic boundary conditions in the
time direction. To address this issue, we average over the different
components corresponding to the same RFs, for instance $\ZA$ is
defined as:
\be
\ZA \equiv \frac{1}{4} \left(\ZA^0+\ZA^1+\ZA^3+\ZA^4 \right)
\ee
where the upper index indicates the Dirac matrix used as current
insertion ($\ZA^i$ corresponds to insertion $\gamma_5\,\gamma_i$). In
addition, remaining systematics are automatically
removed by the subtraction of the ${\cal O}(g^2\,a^\infty)$ terms.
The largest systematic error comes from the choice of the momentum
range to use for the extrapolation to $a^2p^2=0$. One way to estimate
this systematic error is to vary the lower or/and upper range used in
the fit. Another approach is to fix a range and then eliminate a given
momentum in the fit range and refit. The spread of the results about
the mean gives an estimate of the systematic error. In the final
results we give as systematic error the largest one from using these
two procedures, which is the one obtained by modifying the fit range.

\FloatBarrier

Fig.~\ref{ZavALL} corroborates that the magnitude of the lattice
artifacts depends not only on the action parameters, but also on the
operator under study, as can be seen for $\ZA$ and $\ZV$ shown for
different values of the coupling constant. Since both $\ZV$ and $\ZA$
are scale independent, one expects a flat behavior as a function of the
renormalization scale, $(a\,\mu)^2=(a\,p)^2$. However, the
non-perturbative data before subtraction of the lattice artifacts is
carried out, exhibit a non-zero slope, which becomes negligible once
the ${\cal O}(g^2\,a^\infty)$ terms are subtracted. For a proper
comparison, we have kept the y-axis the same as the lattice spacing is
increased. In the case of $\ZA$ we find that the ${\cal O}(g^2\,a^2)$
terms computed for all $N_f{=}4$ gauge configurations, despite being
the leading contributions, underestimate the total one-loop lattice
artifacts, ${\cal O}(g^2\,a^\infty$). Our analysis shows non-negligible
lattice artifacts between $3-6\%$ for momenta in the range
[2,4]. Nevertheless, for the $N_f{=}2$ case, the total one-loop lattice
artifacts are very small ($0.1-2\%$ for $(a\,p)^2: [2,4]$) which may
be attributed to the inclusion of the clover term. One also observes
that the ${\cal O}(g^2\,a^2)$ terms are no longer reliable, possibly
due to the fact that they are polynomial functions of $\csw$ ($2-8\%$
for the aforementioned momentum range). This fact is an evidence that
the addition of the clover term in the twisted mass action suppresses
lattice artifacts. This is also observed for other quantities besides
renormalization functions, such as in the isospin splitting in the
$\Delta$-system~\cite{Alexandrou:2014wca,Alexandrou:2014hsa}. 

For $\ZV$, on the other hand, we find that for all ensembles analyzed
in this work, there are negligible one-loop artifacts beyond
${\cal O}(g^2\,a^2)$, as can be seen in the lower panel of
Fig.~\ref{ZavALL}. From our study we find that lattice artifacs
are current-dependent and can be identified {\it a posteriori}, from
the perturbative computation. This is also confirmed by examining the
results using the $N_f{=}2$ ensembles with the clover term shown in
Figs.~\ref{Zq} - \ref{ZtD}, where for $\ZP$, the ${\cal
  O}(g^2\,a^2)$ and ${\cal O}(g^2\,a^\infty)$ are almost
equivalent, especially for $(a\,p)^2 < 5$; This is not the case for
the other RFs shown in Figs.~\ref{Zq} - \ref{ZtD}.
The $N_f{=}2$ are the most recent gauge configurations produced by
ETMC, which are currently  being used for hadron structure studies and
thus the values of the RFs are needed to renormalize the hadron 
observables~\cite{Abdel-Rehim:2015owa}. Figs.~\ref{Zq} -\ref{ZtD}
correspond to the RFs upon conversion to the ${\overline{\rm MS}}$
scheme at 2 GeV and are plotted against the initial renormalization
scale, $(a\,p)^2$.

\vspace*{0.1cm}
\FloatBarrier

\FloatBarrier

\FloatBarrier

\vspace{0.5cm}
\FloatBarrier

\vspace{0.5cm}
\FloatBarrier

\FloatBarrier

\FloatBarrier

Comparing, for instance, $\ZV$ and $\ZA$ computed using the $N_f{=}4$
and $N_f{=}2$ ensembles (see Fig.~\ref{ZavALL}), we find that the
${\cal O}(g^2\,a^\infty)$ lattice artifacts for the $N_f{=}2$
ensembles are smaller and lead to good quality plateaus. The
extrapolation to $(a\,p)^2 \to 0$  is performed using the 
${\cal O}(g^2\,a^\infty)$-subtracted data and for momenta with
$(a\,p)^2>2$ which is the range of interest. The data display a very
small slope thus leading to a good determination of the continuum
value. For $\ZP$ and $\ZP/\ZS$ there is a stronger dependence on
$(a\,p)^2$ up to $(a\,p)^2\sim 2-3$. Thus, for $\ZP$ and $\ZP/\ZS$ we
use a different interval for the $(a\,p)^2 \to 0$ fit. For instance,
in the $N_f{=}2$ ensembles plotted here, we fit in the range [4,7] for
$\ZP$ and $\ZP/\ZS$, and in the range [2,7] for the remaining RFs. The
systematic errors due to the choice of the fit, are computed by taking
the difference in the values of $Z_{\cal O}^{(0)}$ (see
Eq.~(\ref{Zfinal})) extracted from these ranges and the range [3,5].

\begin{figure}[!h]
{\includegraphics[scale=0.3]{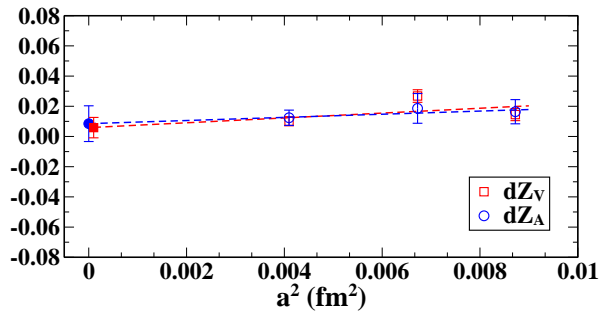}}
\vspace{-0.15cm}
\caption{Difference of $\ZV$ and $\ZA$ computed in this work and in
  Ref.~\cite{Carrasco:2014cwa} (Method 1) for the $N_f{=}4$
  case. The data are plotted against the lattice spacing, and their
  extrapolation to the continuum limit is shown with a dashed
  line. Solid points correspond to the limit $a\to 0$; they are
  consistent with zero.}
\label{dZvdZa} 
\end{figure}

In Table~\ref{tab3} we give our final chiral extrapolated values of
$Z_{\cal O}^{(0)}$ from ${\cal O}(g^2\,a^\infty)$-subtracted data
(e.g. for $N_f{=}2$ the filled diamond point in Figs.~\ref{ZavALL} -
\ref{ZtD}). $\ZP$ and $\ZP/\ZS$ are obtained only at $\beta{=}2.10$
(for both $N_f{=}2,4$) where  we have enough ensembles for the
pion-pole subtraction; the corresponding results are presented in
Table \ref{tab4}. Some of these RFs have been computed in
Ref.~\cite{Carrasco:2014cwa}; the  small differences observed between
our results and those of Ref.~\cite{Carrasco:2014cwa} can be
attributed to two factors: a) the different method for calculating
non-perturbatively the vertex functions (see footnote 1) and b) the
different analysis procedure since the authors use the subtraction of
${\cal O}(g_{boosted}^2\, a^2)$ lattice artifacts, which are, in
general, larger than the ${\cal O}(g^2\,a^\infty)$ terms, 
leading, in some cases, to lower estimates. We have checked that
if we apply the ${\cal O}(g_{boosted}^2\, a^2)$-subtraction to our
non-perturbative estimates, we are in agreement with the
results of Ref.~\cite{Carrasco:2014cwa}. Since the difference between
our approach and that of Ref.~\cite{Carrasco:2014cwa} is the treatment
of lattice artifacts, both sets of results should agree in the
continuum limit, $a\to 0$. This is indeed the case, as demonstrated in
Fig.~\ref{dZvdZa} for the vector and axial RFs. Our results are also
compared to those of Ref.~\cite{Blossier:2014kta}.
We find that upon taking the continuum limit, the differences between
the two works become very small, or compatible with zero. We note that
in Ref.~\cite{Blossier:2014kta} the subtraction of lattice artifacts
is performed via a hypercubic removal procedure. In addition, they use
general momenta that are not restricted to democratic or near democtratic.

\begin{table}[!h]
\begin{center}
\begin{tabular}{cllll}
\hline
\hline
RFs & $\quad N_f{=}2, \beta{=}2.10$ & $\quad N_f{=}4, \beta{=}2.10$ & $\quad N_f{=}4, \beta{=}1.95$ & $\quad N_f{=}4, \beta{=}1.90$  \\[0.21ex]
\hline
\\[-2.5ex]
$\Zq^{\overline{\rm MS}}$   &$\quad$0.8366(2)(7)    $\quad$ &$\quad$0.7822(4)(4)   $\quad$ &$\quad$0.7835(2)(25)  $\quad$ &$\quad$0.7480(6)(11)  $\quad$\\[0.21ex]
$\ZS^{\overline{\rm MS}}$   &$\quad$0.6606(9)(18)   $\quad$ &$\quad$0.7143(9)(216) $\quad$ &$\quad$0.7342(1)(21)  $\quad$ &$\quad$0.7835(2)(17)  $\quad$\\[0.21ex]
$\ZV$                       &$\quad$0.7565(4)(19)   $\quad$ &$\quad$0.6651(2)(5)   $\quad$ &$\quad$0.6298(5)(29)  $\quad$ &$\quad$0.6015(2)(4)   $\quad$\\[0.21ex]
$\ZA$                       &$\quad$0.7910(4)(5)    $\quad$ &$\quad$0.7744(7)(31)  $\quad$ &$\quad$0.7556(5)(85)  $\quad$ &$\quad$0.7474(6)(4)   $\quad$\\[0.21ex]
$\ZT^{\overline{\rm MS}}$   &$\quad$0.8551(2)(15)   $\quad$ &$\quad$0.7875(9)(15)  $\quad$ &$\quad$0.7483(6)(94)  $\quad$ &$\quad$0.7154(6)(6)   $\quad$\\[0.21ex]
$\ZDVa^{\overline{\rm MS}}$ &$\quad$1.1251(27)(17)  $\quad$ &$\quad$1.0991(29)(55) $\quad$ &$\quad$1.0624(108)(33)$\quad$ &$\quad$1.0268(26)(103)$\quad$\\[0.21ex]
$\ZDVb^{\overline{\rm MS}}$ &$\quad$1.1396(16)(13)  $\quad$ &$\quad$1.1398(37)(91) $\quad$ &$\quad$1.1209(61)(32) $\quad$ &$\quad$1.0676(44)(190)$\quad$\\[0.21ex]
$\ZDAa^{\overline{\rm MS}}$ &$\quad$1.1494(9)(99)   $\quad$ &$\quad$1.1741(42)(173)$\quad$ &$\quad$1.1255(27)(328)$\quad$ &$\quad$1.1151(51)(197)$\quad$\\[0.21ex]
$\ZDAb^{\overline{\rm MS}}$ &$\quad$1.1357(20)(205) $\quad$ &$\quad$1.1819(47)(147)$\quad$ &$\quad$1.1555(36)(289)$\quad$ &$\quad$1.1170(54)(223)$\quad$\\[0.21ex]
$\ZDTa^{\overline{\rm MS}}$ &$\quad$1.1377(160)(13) $\quad$ &$\quad$1.1562(32)(7)  $\quad$ &$\quad$1.1218(106)(44)$\quad$ &$\quad$1.0777(37)(122)$\quad$\\[0.21ex]
$\ZDTb^{\overline{\rm MS}}$ &$\quad$1.1472(121)(48) $\quad$ &$\quad$1.1822(59)(118)$\quad$ &$\quad$1.1727(121)(73)$\quad$ &$\quad$1.0965(90)(278)$\quad$\\[0.2ex]
\hline
\hline
\end{tabular}
\caption{Our final values of the renormalization functions. The
scheme and scale dependent RFs are given in ${\overline{\rm MS}}$
at 2 GeV. The number in the first parenthesis is the statistical error,
while the number in the second parenthesis corresponds to the
systematic error obtained by varying the fit range in the
$(a\,p)^2\to 0$ extrapolation.} 
\label{tab3}
\end{center}
\end{table}
\FloatBarrier
\vspace*{-0.25cm}
\begin{table}[!h]
\begin{center}
\begin{tabular}{cll}
\hline
\hline
RFs & $\quad N_f{=}2, \beta{=}2.10$ & $\quad N_f{=}4, \beta{=}2.10$  \\
\hline
\\[-2.5ex]
$\ZP^{\overline{\rm MS}}$   &$\quad$0.5012(75)(258) $\quad$ &$\quad$0.5468(15)(176)$\quad$ \\[0.21ex]
$\ZP/\ZS$                   &$\quad$0.7016(141)(113)$\quad$ &$\quad$0.7036(23)(195)$\quad$ \\[0.2ex]
\hline
\hline
\end{tabular}
\caption{Our final values for $\ZP^{\overline{\rm MS}}(2\,{\rm GeV})$ 
and $\ZP/\ZS$. The
number in the first parenthesis is the statistical error, while the
number in the second parenthesis corresponds to the systematic error
obtained by varying the fit range in the $(a\,p)^2\to 0$ extrapolation.} 
\label{tab4}
\end{center}
\end{table}
\FloatBarrier
\vspace*{-0.25cm}
\FloatBarrier

\section{Conclusions}
\label{sec6}

We present results on the renormalization functions of the
fermion field and fermion bilinears with up to one covariant
derivative. The computation is performed non-perturbatively on
several ensembles of $N_f{=}4$ twisted mass fermions, as well as
$N_f{=}2$ twisted mass fermions including a clover term.
This work is a continuation of our renormalization program first
addressed in Refs.~\cite{Alexandrou:2010me,Alexandrou:2012mt}. Besides
the analysis of the $N_f{=}2$ twisted mass clover-improved ensembles 
we have improved the procedure for the subtraction of lattice
artifacts. The procedure that we adopt here for the perturbative
computation of lattice artifacts is improved by taking into account
not only leading order lattice artifacts, ${\cal O}(g^2\,a^2)$, but
also contributions to all orders in the lattice spacing, 
${\cal O}(g^2\,a^\infty)$. 

The non-perturbative computation uses a momentum-dependent source and
the RFs are extracted for all the relevant operators
simultaneously. This leads to a very accurate evaluation of the RFs
using only a small ensemble of gauge configurations (${\cal O}(10)$). 
The precision of the results allows us to reliably investigate the
quark mass dependence, which is found to be very weak with the
exception of $\ZP$. Nevertheless, a linear extrapolation with
respect to the pion mass squared is carried out in order to reach the
chiral limit. For the renormalization function of the pseudoscalar
operator, $\ZP$, we find a quark mass dependence due to the pion pole,
which we eliminate  using a refined procedure that avoids introduction
of artificially large errors. The procedure entails  a suitable fit
(Eq.~(\ref{PionPole})) to identify $\ZP$ directly from the constant
term, $a_P$, instead of subtracting the pion-pole term (see
Eq.~(\ref{polesub})).

Our accurate non-perturbative results show that, although the lattice
spacings considered in this work are smaller than $0.1$~fm, lattice
artifacts are not negligible in most cases, and are significantly
larger than statistical errors. Thus, the subtraction of the 
${\cal O}(g^2\,a^\infty)$ perturbative contributions appear to improve
significantly the determination of the RFs, by leading to a milder
dependence of the RFs on $(ap)^2$. Residual ${\cal O}(a^2p^2)$ effects
are removed by extrapolating our results to $(ap)^2=0$. For the scheme
and scale dependent RFs, we convert our values to the $\rm \overline{MS}$ 
scheme at a scale of 2~GeV. The statistical errors are, in general,
smaller than the systematic ones. The latter are estimated by changing
the window of values of the momentum used to extrapolate to $a^2p^2=0$. 
Our final values are given in Tables~\ref{tab3} - \ref{tab4}.

\section{Acknowledgments}

We would like to thank all members of ETMC for a
very constructive and enjoyable collaboration and for the fruitful
discussions. This work used computational resources provided by PRACE, JSC,
Germany as well as by the Cy-Tera machine at the Cyprus Institute. This
work is in part supported by funding received from the Cyprus Research
Promotion Foundation under contract NEA Y$\Pi$O$\Delta$OMH/$\Sigma$TPATH/0308/31 
co-financed by the European Regional Development Fund. M.C. acknowledges 
financial support received from the Cyprus Research Promotion
Foundation under contract TECHNOLOGY/$\Theta$E$\Pi$I$\Sigma$/0311(BE)/16.

\appendix

\section{$\beta-$function and anomalous dimensions}
\label{appA}

For completeness we  provide in this Appendix the definition of
the $\beta-$function and the anomalous dimension of the operators
studied in this work, which include up to one covariant derivative. To
simplify the expressions we give the perturbative coefficients in the
Landau gauge and in $SU(3)$.

The perturbative expansion of the anomalous dimension in a
renormalization scheme $\mathcal S$ is given as follows:
\be
\gamma^{\mathcal S} = - \mu \frac{\mathrm d}{\mathrm d \mu}
\log Z_{\mathcal S} =
\gamma_0 \frac{g^{\mathcal S} (\mu)^2}{16 \pi^2}
 + \gamma_1^{\mathcal S}
       \left( \frac{g^{\mathcal S} (\mu)^2}{16 \pi^2} \right)^2
 + \gamma_2^{\mathcal S}
       \left( \frac{g^{\mathcal S} (\mu)^2}{16 \pi^2} \right)^3
 + \cdots
\label{gammaS}
\ee
Similarily, the $\beta-$function is defined as:
\be
\beta^{\mathcal S} =  \mu \frac{\mathrm d}{\mathrm d \mu}
                      g^{\mathcal S} (\mu) =
 - \beta_0 \frac{g^{\mathcal S} (\mu)^3}{16 \pi^2}
 - \beta_1 \frac{g^{\mathcal S} (\mu)^5}{(16 \pi^2)^2}
 - \beta_2^{\mathcal S} \frac{g^{\mathcal S} (\mu)^7}{(16 \pi^2)^3}
 + \cdots\,.
\ee
For the conversion from the RI$'$ to the ${\overline{\rm MS}}$ scheme
we use the three-loop expressions, to which the coefficients of the
$\beta-$function coincide and are given by~\cite{vanRitbergen:1997va,Gracey:2003yr}:
\bea
\beta_0 & = & 11 - \frac{2}{3} N_f \,, \\
\beta_1 & = & 102 - \frac{38}{3} N_f \,, \\
\beta_2 & = & \frac{2857}{2} - \frac{5033}{18} N_f
              + \frac{325}{54} N_f^2 \,.
\eea
Below we give all necessary expressions to convert to the
${\overline{\rm MS}}$ scheme, as well as the references from which
they were taken (see also references therein). Dome signs and 
multiplicative numerical factors have been adjusted to match the
definition of Eq.~(\ref{gammaS}). An upper index appears for
scheme-dependent quantities, in order to denote the scheme that they
correspond to.

{\underline{Quark field}}~\cite{Chetyrkin:1999pq}:
\bea
\gamma_0 & = & 0 \,, \\
\gamma_1 & = & \frac{134}{3} - \frac{8}{3} N_f \,, \\
\gamma_2^{\overline{\rm MS}} & = & \frac{20729}{18} - 79 \zeta_3 - \frac{1100}{9} N_f
               + \frac{40}{27} N_f^2 \,,\\
\gamma_2^{{\mbox{\scriptsize RI$^{\prime}$}}} & = & \frac{52321}{18} - 79 \zeta_3 - \frac{1100}{9}
N_f + \frac{40}{27} N_f^2\,,
\eea
($\zeta_3 = 1.20206...$) 

{\underline{Scalar/pseudoscalar}}~\cite{Chetyrkin:1997dh,Vermaseren:1997fq}:
\bea
\gamma_0 & = & -8 \,, \\
\gamma_1^{\overline{\rm MS}} & = & - \frac{404}{3} + \frac{40}{9} N_f \,, \\
\gamma_1^{\mbox{\scriptsize RI$^{\prime}$}} & = & - 252 + \frac{104}{9} N_f \,, \\
\gamma_2^{\overline{\rm MS}} & = & - 2498 + \left( \frac{4432}{27}
           + \frac{320}{3} \zeta_3 \right) N_f + \frac{280}{81} N_f^2 \,,\\
\gamma_2^{\mbox{\scriptsize RI$^{\prime}$}} & = & - \frac{40348}{3} + \frac{6688}{3}\zeta_3 +\left( \frac{35176}{27}
           - \frac{256}{9} \zeta_3 \right) N_f - \frac{1712}{81} N_f^2 \,,
\eea

{\underline{Tensor}}~\cite{Gracey:2000am,Gracey:2003yr}:
\bea
\gamma_0 & = & \frac{8}{3} \,, \\
\gamma_1 & = & \frac{724}{9} - \frac{104}{27} N_f \,, \\
\gamma_2^{\overline{\rm MS}} & = & \frac{105110}{81} - \frac{1856}{27} \zeta_3
             - \left( \frac{10480}{81}
             + \frac{320}{9} \zeta_3 \right) N_f - \frac{8}{9} N_f^2\,,\\
\gamma_2^{\mbox{\scriptsize RI$^{\prime}$}} & = & \frac{359012}{81} - \frac{26144}{27} \zeta_3
             + \left(-\frac{39640}{81}
             + \frac{512}{27} \zeta_3 \right) N_f + \frac{2288}{243} N_f^2\,.
\eea

{\underline{One-derivative vector/axial}}~\cite{Gracey:2003mr,Gockeler:2010yr}:
\bea
\gamma_0 & = & \frac{64}{9} \,, \\
\gamma_1^{\overline{\rm MS}} & = & \frac{23488}{243} - \frac{512}{81} N_f \,, \\
\gamma_1^{\mbox{\scriptsize RI$^{\prime}$}} & = & \frac{48040}{243} - \frac{112}{9} N_f \,, \\
\gamma_2^{\overline{\rm MS}} & = & \frac{11028416}{6561} + \frac{2560}{81} \zeta_3
             - \left( \frac{334400}{2187}
             + \frac{2560}{27} \zeta_3 \right) N_f - \frac{1792}{729} N_f^2\,,\\
\gamma_2^{\mbox{\scriptsize RI$^{\prime}$}} & = & \frac{59056304}{6561} 
- \frac{103568}{81} \zeta_3 - \left(\frac{2491456}{2187}
+ \frac{416}{27} \zeta_3 \right) N_f + \frac{19552}{729} N_f^2\,.
\eea

{\underline{One-derivative tensor}}~\cite{Gracey:2003mr}:
\bea
\gamma_0 & = & 8\,, \\
\gamma_1^{\overline{\rm MS}} & = & 124 - 8\,N_f \,, \\
\gamma_1^{\mbox{\scriptsize RI$^{\prime}$}} & = & \frac{680}{3} - \frac{128}{9} N_f \,, \\
\gamma_2^{\overline{\rm MS}} & = & \frac{19162}{9} 
             - \left( \frac{5608}{27}
             + \frac{320}{3} \zeta_3 \right) N_f - \frac{184}{81} N_f^2\,,\\
\gamma_2^{\mbox{\scriptsize RI$^{\prime}$}} & = & \frac{97052}{9} 
- \frac{4312}{3} \zeta_3 - \left(\frac{36848}{27}
+ \frac{176}{9} \zeta_3 \right) N_f + \frac{2624}{81} N_f^2\,.
\eea

\section{Application of the subtraction to ${\cal O}(g^2\,a^\infty)$ in other ensembles}
\label{appB}

As discussed in the main part of the paper in our previous works of Refs.
\cite{Alexandrou:2010me,Alexandrou:2012mt} we have applied a procedure
of subtracting the lattice artifacts of ${\cal O}(g^2\,a^2)$. The
values of the RFs are used to renormalize hadron quantities such
as the axial charge and the quark momentum fraction, in order to
compare them with other lattice discretizations, as well as with
experimental data. For a fair comparison between renormalized matrix
elements of the ensembles presented in this work and the ones given in
Refs.~\cite{Alexandrou:2010me,Alexandrou:2012mt}, we have updated the
RFs of the latter publications by applying the subtraction procedure
to one-loop and all orders in the lattice spacing, ${\cal O}(g^2\,a^\infty)$.
These correspond to tree-level Symanzik improved gauge action and
$N_f{=}2$ twisted mass fermions at three values of the
coupling constant corresponding to $\beta{=}3.90,\,4.05,\,4.20$. Since
the gluon action is different from the ensembles of Table~\ref{tab1},
and since employed momentum values are also different, a perturbative
computation of the ${\cal O}(g^2\,a^\infty)$ contributions was
required on each ensemble in order to match its parameters, such as
the coupling constant, the lattice size and the values of the
renormalization scales. The new data on the Rfs are given in
Table~\ref{tab5}.

\begin{table}[!h]
\begin{center}
\begin{tabular}{clll}
\hline
\hline
$\qquad$RFs$\qquad$ & $\quad$$\beta=3.9$$\qquad$  &  $\quad$$\beta=4.05$$\qquad$  &   $\quad$$\beta=4.20$$\qquad$     \\
\hline
\\[-2.5ex]
$\Zq^{\overline{\rm MS}}$      &0.769(1)(2)   &0.787(1)(3)   &0.783(1)(2)    \\[0.21ex]
$\ZS^{\overline{\rm MS}}$      &0.791(2)(41)  &0.748(2)(31)  &0.754(1)(16)   \\[0.21ex]
$\ZP^{\overline{\rm MS}}$      &0.527(6)(70)  &0.517(2)(33)  &0.546(5)(33)   \\[0.21ex]
$\ZP/\ZS$                      &0.672(7)(60)  &0.700(3)(14)  &0.731(5)(25)   \\[0.21ex]
$\ZV$                          &0.646(2)(2)   &0.681(2)(6)   &0.701(1)(4)    \\[0.21ex]
$\ZA$                          &0.769(2)(1)   &0.787(1)(1)   &0.791(1)(1)    \\[0.21ex]
$\ZT^{\overline{\rm MS}}$      &0.758(2)(4)   &0.796(1)(3)   &0.814(1)(3)    \\[0.21ex]
$\ZDVa^{\overline{\rm MS}}$    &1.028(2)(6)   &1.080(2)(11)  &1.087(3)(12)    \\[0.21ex]
$\ZDVb^{\overline{\rm MS}}$    &1.064(4)(4)   &1.123(4)(10)  &1.130(4)(4)    \\[0.21ex]
$\ZDAa^{\overline{\rm MS}}$    &1.106(3)(8)   &1.157(4)(10)  &1.150(4)(15)   \\[0.21ex]
$\ZDAb^{\overline{\rm MS}}$    &1.102(5)(7)   &1.161(4)(13)  &1.164(3)(6)    \\[0.2ex]
\hline
\hline
\end{tabular}
\caption{Updated results on the RFs of
Refs.~\cite{Alexandrou:2010me,Alexandrou:2012mt} ($N_f{=}2$,
$\beta{=}3.90,\,4.05,\,4.20$, $\csw=0$) using the subtraction procedure to
${\cal O}(g^2\,a^\infty)$.}
\label{tab5}
\end{center}
\end{table}
\FloatBarrier

\bibliographystyle{apsrev}                     % Style for bibliography
\bibliography{Zfactors_Nf4_Nf2csw}

\end{document}